\documentclass[preprint]{elsarticle}

\usepackage{lineno,hyperref}
\modulolinenumbers[5]

\journal{Journal of \LaTeX\ Templates}

\usepackage{graphicx}
\usepackage{multirow}
\usepackage{amssymb}
\usepackage{wrapfig} 
\usepackage{enumerate}
\usepackage[shortlabels]{enumitem}
\usepackage{todonotes}
\usepackage{pifont}
\usepackage{booktabs} 
\usepackage{longtable}
\usepackage{lscape}
\usepackage{verbatim} 
\usepackage{todonotes}
\usepackage{comment}
\usepackage{enumerate}

\begin{document}

\begin{frontmatter}

\title{Motivating the Contributions: An Open Innovation Perspective on What to Share as Open Source Software}

\author[LU]{J.~Lin{\aa}ker\corref{corrAuthor}}
\ead{johan.linaker@cs.lth.se}
\author[LU]{H.~Munir}
\ead{hussan.munir@cs.lth.se}
\author[BTH]{K.~Wnuk}
\ead{krw@bth.se}
\author[Sony]{C.E.~Mols}
\ead{carl-eric.mols@sonymobile.com}

\cortext[corrAuthor]{Corresponding author}

\address[LU]{Lund University, Box 118, 221 00 Lund, Sweden}
\address[BTH]{Blekinge Institute of Technology, 371 79 Karlskrona, Sweden}
\address[Sony]{Sony Mobile, Mobilv{\"a}gen 10, Lund, Sweden}

\begin{abstract}
Open Source Software (OSS) ecosystems have reshaped the ways how software-intensive firms develop products and deliver value to customers. However, firms still need support for strategic product planning in terms of what to develop internally and what to share as OSS. Existing models accurately capture commoditization in software business, but lack operational support to decide what contribution strategy to employ in terms of what and when to contribute. This study proposes a Contribution Acceptance Process (CAP) model from which firms can adopt contribution strategies that align with product strategies and planning. In a design science influenced case study executed at Sony Mobile, the CAP model was iteratively developed in close collaboration with the firm's practitioners. The CAP model helps classify artifacts according to business impact and control complexity so firms may estimate and plan whether an artifact should be contributed or not. Further, an information meta-model is proposed that helps operationalize the CAP model at the organization. The CAP model provides an operational OI perspective on what firms involved in OSS ecosystems should share, by helping them motivate contributions through the creation of contribution strategies. The goal is to help maximize return on investment and sustain needed influence in OSS ecosystems.
\end{abstract}


\end{frontmatter}

\section{Introduction}
\label{intro}

Open Innovation (OI) has attracted scholarly interest from a wide range of disciplines since its introduction~\cite{west2013leveraging}, but remains generally unexplored in software engineering~\cite{munir2015open}. A notable exception is that of Open Source Software (OSS) ecosystems~\cite{jansen2009business, west2003open, west2006challenges}. Directly or indirectly adopting OSS as part of a firm's business model~\cite{chesbrough2007open} may help the firm to accelerate its internal innovation process~\cite{chesbrough2003open}. One reason for this lies in the access to an external workforce, which may imply that costs can be reduced due to lower internal maintenance and higher product quality, as well as a faster time-to-market~\cite{stuermer2009extending, ven2008challenges}. A further potential benefit is the inflow of features from the OSS ecosystem. This phenomenon is explained by Joy's law as \textit{``no matter who you are, not all smart people work for you''}. 

From an industry perspective, these benefits are highlighted in a recent study of 489 projects from European organizations that showed projects of organizations involving OI achieved a better financial return on investment compared to organizations that did not involve OI \cite{du2014research}. Further, two other studies \cite{laursen2006open,muniropen2016} have shown that organizations with more sources of external knowledge achieved better product and process innovation for organization's proprietary products. Moreover, a recent survey study~\cite{chesbrough2014fad} in 125 large firms of EU and US showed that 78\% of organizations in the survey are practicing OI and neither of them has abandoned it since the introduction of OI in the organization. This intense practicing of OI also leads 82\% of the organizations to increase management support for it and 53\% of the organizations to designate more than 5 employees working full-time with OI. Moreover, the evidence suggests that 61\% of the organizations have increased the financial investment and 22\% have increased the financial investment by 50\% in OI.

To better realize the potential benefits of OI resulting from participation in OSS ecosystems, firms need to establish synchronization mechanisms between their product strategy and product planning~\cite{fricker2012software}, and how they participate in the ecosystems and position themselves in the ecosystem governance structures~\cite{munir2015open, wnuk2012can, stam2009when, baars2012framework}. This primarily concerns firms that either base their products on OSS or employ OSS as part of their sourcing strategy. To achieve this synchronization, these firms need to enrich their product planning and definition activities with a strategic perspective that involves what to keep closed and what to contribute as OSS. We label this type of synchronization as \textit{strategic product planning} in OI. \textit{Contribution strategies}~\cite{wnuk2012can}, i.e., guidelines that explain \textit{what} should be contributed, and \textit{when} play a vital role here. A common strategy is to contribute parts considered as a commodity while keeping differentiating parts closed~\cite{west2003open, henkel2006selective}. The timing aspect is critical as functionality sooner or later will pass over from being differentiating to commodity due to a constantly progressing technology life-cycle~\cite{van2009commodification}. This strategy is further emphasized by existing commoditization models~\cite{van2009commodification, Bosch13}. However, these models are not designed with active OSS ecosystem participation in mind and lack support for strategic product planning and contribution strategies.

In this paper, we occupy this research gap by presenting a Contribution Acceptance Process (CAP) model. The model was developed in close collaboration with Sony Mobile. Sony Mobile is actively involved in a number of OSS ecosystem, both in regard to their products features and their internal development infrastructure\footnote{http://developer.sonymobile.com/knowledge-base/open-source/}. With the consideration of OSS as an external asset, the CAP model is based on the Kraljic's portfolio purchasing model which helps firms analyze risk and maximize profit when sourcing material for their product manufacturing~\cite{kraljic1983purchasing}. The original model is adapted through an extensive investigation of Sony Mobile's contribution processes and policies, and designed to support firms' strategic product planning. More specifically, the model helps firms to create contribution strategies for their products and software artifacts such as features and components. Hence, the CAP model is an important step for firms that use OSS ecosystems in their product development and want to gain or increase the OI benefits, such as increased innovation and reduced time-to-market. Moreover, we help firms to operationalize the CAP model by proposing an information meta-model. The meta-model is an information support that should be integrated into the requirements management infrastructure and enables contribution strategies to be communicated and followed up on a software artifact-level throughout a firm's development organization. As a first validation outside of Sony Mobile, the CAP model was presented to and applied in three case firms. This provided understanding of the model's generalizability, and also input to future design cycles.



The rest of the paper is structured as follows: In section~\ref{sec:RW}, we position our study with related work and further motivate the underlying research gap. This is followed by section~\ref{sec:Design} in which we describe the research design of our study, its threats to validity and strategies used to minimize these threats. In section~\ref{sec:Model} we present our CAP model and in section~\ref{sec:Repositories} we present an information meta-model for how contribution decisions may be traced. In section~\ref{sec:Example}, we present an example of how the CAP model and meta-model may be used together inside Sony Mobile. In section~\ref{sec:CaseStudies} we present findings from three exploratory case studies outside Sony Mobile where we focused on early validation the CAP model's applicability and usability. Finally, in section~\ref{sec:Discussion} we discuss the CAP model in relation to related work, and specific considerations, while we summarize our study in section~\ref{sec:Conclusions}.

\section{Related Work}\label{sec:RW}
Below we describe the context of our research with respect to how software engineering and OSS fits into the context of OI. Further, we give a background on contribution strategies and commoditization models. Moreover, we provide a background of the sourcing model on which the CAP model is based. We than provide an overview on what we label as strategic product planning, as well as on software artifacts, and conclude by describing the research gap, that this study aims to fill.

\subsection{Open Innovation in Software Engineering}
\begin{figure}
\begin{center}
\includegraphics[scale=0.55]{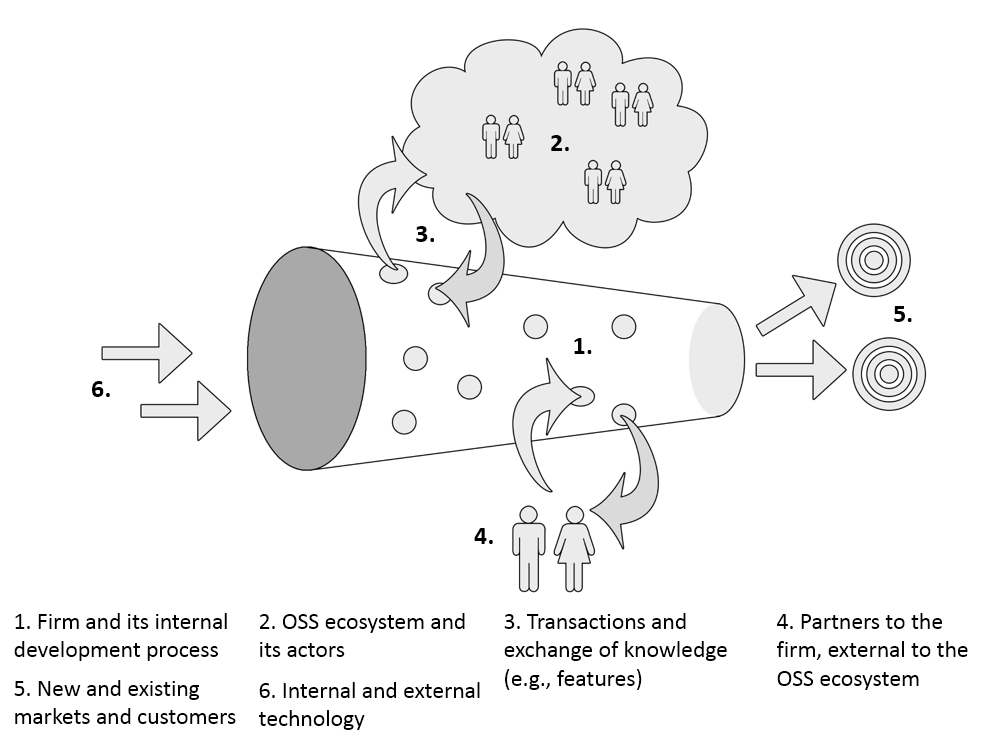}
\caption{The OI model illustrated with interactions between the firm (1) and its external collaborations (2,4). Adopted from Chesbrough~\cite{chesbrough2006open}.}
\label{fig:Funnel}
\end{center}
\end{figure}

OI is commonly explained by a funnel model~\cite{chesbrough2006open} representing a firm's R\&D process, see Fig.~\ref{fig:Funnel}. The funnel (1) is permeable, meaning that the firm can interact with the open environment surrounding it. This conceptualization fits onto many contexts, e.g., a firm that takes part in a joint-venture or start-up acquisition. In our case, we focus on ecosystems (2) and specifically those based on OSS~\cite{jansen2009business, garcia2017preface}. An OSS ecosystem consists of the focal firm along with other actors who jointly see to the development and maintenance of an OSS project, which may be seen as the technological platform underpinning the relationships between the actors~\cite{jansen2009sense, manikas2013software}. In the context of this study, the focal firm represented by the OI funnel is Sony Mobile and their internal software development process. The OSS ecosystem could, for example, be represented by that surrounding the Android Open Source Project\footnote{https://source.android.com/} (AOSP). 
The interactions between the focal firm and the ecosystem (see Fig.~\ref{fig:Funnel}) are represented by the arrows going in and out and can be further characterized as knowledge exchange between the firm and the OSS ecosystem (e.g., Sony Mobile and AOSP). Examples of transactions can include software artifacts (e.g., bug fixes, features, plug-ins, or complete projects), but also opinions, knowledge, and support that could affect any step of the internal or external development. 

The interactions (3) may be bi-directional in the sense that they can go into the development process from the open environment (\textit{outside-in}), or from the development process out to the open environment (\textit{inside-out}). \textit{Coupled innovation}~\cite{enkel2009open} happens when outside-in and inside-out transactions occurs together (i.e., consumption of and contribution to OSS). This may be expected in co-development between a firm and other ecosystem participants in regard to specific functionality (e.g., Sony Mobile's developer toolkits\footnote{https://github.com/sonyxperiadev}).

How firms choose to work with and leverage these interactions with OSS ecosystems impact how they will realize the potential benefits of OI, such as increased innovation, shorter time-to-market, and better resource allocation~\cite{stuermer2009extending, ven2008challenges}. The CAP model presented in this paper provides operational and decision-making guidelines for these firms in terms what they should contribute to and source of from the OSS ecosystems. I.e., how they should interact with the open environment in an inside-out, outside-in, or coupled direction. Hence, what the CAP model brings in terms of novelty is an operational OI perspective on what firms involved in OSS ecosystems should share, by helping firms motivate the contributions through the creation of tailored contribution strategies.


\subsection{Contribution Strategies in Open Source Software Ecosystem}

Wnuk et al.~\cite{wnuk2012can} define a contribution strategy as a managerial practice that helps to decide what to contribute as OSS, and when. 
To know what to contribute, it is important for firms to understand how they participate in various OSS ecosystems in regards to their business model and product strategy from an OI perspective. Dahlander \& Magnusson~\cite{dahlander2008firms} describe how a firm may access the OSS ecosystems in order to extend its resource base and align its product strategy with ecosystems' strategies. In another study, Dahlander \& Magnusson~\cite{dahlander2005relationships} describe how a firm can adapt its relationships with the OSS ecosystems based on how much influence the firm needs, e.g., by openly contributing back to the OSS ecosystem, or by keeping new features internal. To build and regulate these relationships, a firm can apply different revealing strategies in this regard: differentiating parts are kept internal while commodity parts are contributed~\cite{henkel2006selective, west2003open}. Further, licenses may be used so that the technology can be disclosed under conditions where control is still maintained~\cite{west2003open}. Depending on the revealing strategy the level of openness may vary from completely open, partly transparent conditions~\cite{chesbrough2007open}, to completely closed.
As highlighted by Jansen et al.~\cite{jansen2012shades}, the openness of a firm should be considered as a continuum rather than a binary choice.


\subsection{Commoditization Models}
\label{subsec:CommoditizationModels}
With commoditization models, we refer to models that describe a software artifact's value depreciation ~\cite{SMR1560} and how it moves between a differential to a commodity state, i.e., to what extent the artifact is considered to help distinguish the focal firm's product offering relative to its competitors. Such models can help firms better understand what they should contribute to OSS ecosystems, and when, i.e., provide a base to design contribution strategies~\cite{wnuk2012can}. Van der Linden et al.~\cite{van2009commodification} stressed that efficient software development should focus \textit{``\ldots on producing only the differentiating parts''} and that \textit{``\ldots preferably, firms acquire the commodity software elsewhere, through a distributed development and external software such as [commercial software] or OSS''}. Firms should hence set the differentiating value of a software artifact in relation to how it should be developed, or even if it should be acquired. Commoditization is also related to the product's life-cycle and, is more often experienced towards the end of the life cycle ~\cite{kittlaus2008software}. 

Van der Linden et al.~\cite{van2009commodification} present a commoditization model that highlights how commoditization is a continuous and inevitable process for all software artifacts. Therefore, firms should consider whether the software or technology should be developed, acquired, or kept internally, shared with other firms, or made completely open (e.g., as OSS)~\cite{Badampudi2016105}. Ideally, differentiating software or technology should be kept internal, but as their life-cycle progresses their value depreciates and they should be made open. This is particularly relevant for software artifacts that have an enabling role for cross-value creation, data collection or support value creation when combined with other parts of the offering, e.g., an artifact that collects and analyzes anonymous customer data that could be offered as business intelligence to customers~\cite{SMR1560}. Bosch~\cite{Bosch13} presents a similar commoditization model, which classifies the software into three layers and describes how a software's functionality moves from an early development stage as experimental and innovative, to a more mature stage where it provides special value to customers and advantage towards competition, then finally transitioning to stage where it is considered as commodity, hence it \textit{``\ldots no longer adds any real value''}~\cite{Bosch13}.


A challenge identified by both van der Linden et al.~\cite{van2009commodification} and Bosch~\cite{Bosch13} is the risk of losing Intellectual property rights (IPR) to competitors, a challenge that has also been highlighted in numerous other studies ~\cite{wnuk2012can,henkel2006selective,henkel2008champions,west2006challenges}. By not contributing software and technology that are considered differentiating, firms can avoid the risk of giving away its added value to competitors. However, both van der Linden et al.~\cite{van2009commodification} and Bosch~\cite{Bosch13} highlight how the acquisition of the commodity functionality may help firms to reduce the development and maintenance cost, and potentially shorten time-to-market. Instead, they can shift internal focus to differential features and better-justified R\&D activities~\cite{van2009commodification}. 

\subsection{The Kraljic Portfolio Purchasing Model}
\label{subsec:Kraljic}
From the software product planning perspective, sourcing refers to decisions of what parts of the software that should be developed internally or acquired externally, from where and how~\cite{kittlaus2008software}, and is an important part of a firm's product strategy~\cite{fricker2012software}. A recent literature review of software component decision-making making lists four sourcing strategies: in-house, outsourcing, COTS and OSS and brings supporting evidence that two sourcing strategies are often considered~\cite{Badampudi2016105}. From an OSS perspective, sourcing, therefore, regards decisions on if, and what, parts of the internal software that should be based on and/or co-developed as OSS (also referred to as \textit{Open-Sourcing}~\cite{aagerfalk2008outsourcing}). This is further highlighted in existing commoditization models (see section~\ref{subsec:CommoditizationModels}), which argues how commodity parts should be acquired, contributed and sourced in different ways, while internal development should be focused on differenting parts~\cite{van2009commodification, Bosch13}. With this background, we have chosen to base the CAP-model presented in this study on the portfolio purchasing model by Peter Kraljic~\cite{kraljic1983purchasing}. 

\begin{figure*}
\begin{center}
\includegraphics[width=\textwidth]{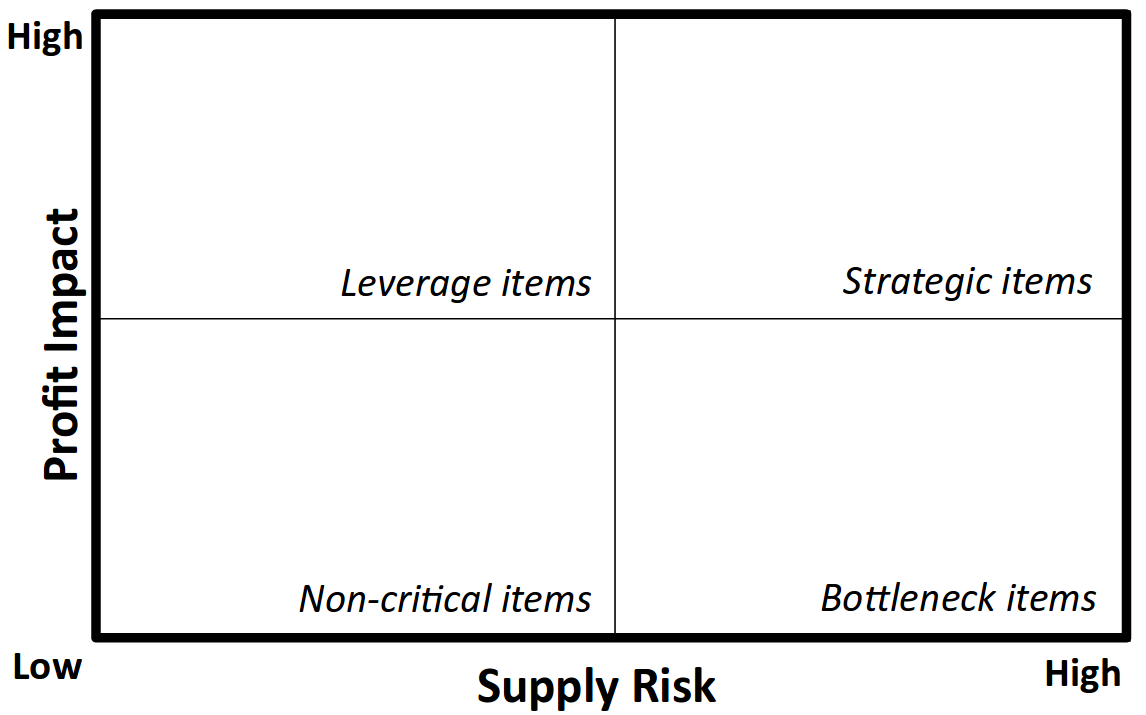}
\label{fig:KraljicMatrix}
\caption{The matrix used in Kraljic's portfolio purchasing model~\cite{kraljic1983purchasing}, which allows supply-items needed for a product to classified into four item categories based on the two dimensions Business impact and Supply risk.}
\end{center}
\end{figure*}

Kraljic's model describes how to develop a sourcing strategy for the supply-items (e.g., material and components) required for a product. First, the supply-items are classified according to the \textit{Profit impact} and \textit{Supply risk} dimensions on a scale from low to high. The profit impact concerns the strategic importance of the item, as well as the added value and costs which that it generates for the firm. The supply risk refers to the availability of the item, ease to substitute its suppliers, and how it is controlled. The supply items are then positioned onto a matrix with four quadrants, based on the two dimensions, see Fig.~\ref{fig:KraljicMatrix}. Each quadrant represents a specific item category with its own distinctive purchasing strategy towards the suppliers~\cite{kraljic1983purchasing}. 

\begin{itemize}
    \item \textbf{Strategic items:} These are items with high-profit impact and high supply risk. They can usually only be acquired from a single supplier. A common strategy is to form and maintain a strategic partnership with the supplier~\cite{Caniels2005141}.
    \item \textbf{Leverage items:} These are items with high-profit impact and low supply risk. Can generally be obtained from multiple suppliers at a low switching cost. A common strategy is to exploit buying power within the supplier market~\cite{Caniels2005141}.
    \item \textbf{Bottleneck items:} These are items with low-profit impact and high supply risk. Suppliers are usually in a dominant position. A common strategy is to accept dependence and strive to reduce negative effects, e.g., through risk analysis and stock-piling~\cite{Caniels2005141}.
    \item \textbf{Non-critical items:} These are items with low-profit impact and low supply risk. They generally have a low added-value per item. A general strategy is to reduce related costs, such as logistic and administrative~\cite{Caniels2005141}.
\end{itemize} 

Determining how a material or component should be classified may be done in several ways. Gelderman et al.~\cite{gelderman2003handling} report how a consensus-seeking method is frequently used by inviting cross-functional competencies and internal stakeholders to discuss how items should be rated in regard to the two dimensions~\cite{gelderman2003handling}. Other measurement approaches involve representing each dimension with a specific variable (e.g., supply risk as a number of available suppliers), or using a set of variable and weighting them together. After a set of items have been analyzed and put on the matrix, discussions, and reflections are performed and can potentially lead to a revision of the item categorization~\cite{gelderman2003handling}. This discussion may concern how the firm should maintain the items' current positions or strive to move certain items between the quadrants.

The model inspired several industries and academics. Among some examples, Cani\"{e}ls and Gelderman~\cite{Caniels2005141} studied the choice of various purchasing strategies and empirically quantified the "relative power" and "total interdependence" aspects among Dutch purchasing professionals. Ulkuniemi et al.~\cite{Ulkuniemi201554} looked at purchasing as a market shaping mechanism and identified five types of market shaping actions. Shaya discussed the usage of the Kraljic's portfolio model for optimizing the process of sourcing IT and managing software licenses at Skanska ITN~\cite{shaya2012process}. Gangadharan et al. proposed using Kraljic's portfolio model for mapping SaaS services and sourcing structure~\cite{Gangadharan2016}. 
To the best of our knowledge, no study has suggested using Kraljic's model in the context of OSS ecosystems and creation of contribution strategies for software artifacts. 

\subsection{Strategic Product Planning in OI}
A software product strategy defines the product and describes how it will evolve for a longer period of time~\cite{fricker2012software}. It should consider aspects such as the product definition in terms of functional and quality scope, target market, delivery model, positioning and sourcing~\footnote{http://community.ispma.org/body-of-knowledge/}. Product planning executes product strategy with the help of roadmapping, release planning, and requirements management processes~\cite{fricker2012software}. Hence, decisions regarding if, and what parts of the product should be based on OSS concerns executive management and the software product management (SPM) as they usually oversee the product strategy~\cite{Maglyas}, but also the development organization as they, together with SPM, oversee the product planning and development. 

To the best of our knowledge, the current literature offers limited operational support for creating contribution strategies that help synchronize product strategies and product planning with OSS ecosystems. Therefore, we present the CAP model to support software firms in building strategic product planning that looks beyond realizing a set of features in a series of software releases that reflects the overall product strategy and adds the strategic OI aspect with the help of contribution strategies.


\subsection{Artifacts in Software Engineering}
The CAP model presented in this paper offers a tool for firms to decide whether or not a software artifact should be contributed to an OSS ecosystem or not. In this context, a software artifact may refer to a functionality of different abstractions, e.g., bug-fixes, requirements, features, architectural assets or components. These artifacts may be represented and linked together in software artifact repositories~\cite{mendezfernandez2012field}, often used for gathering, specification and communication of requirements inside a software development organization's requirements management infrastructure~\cite{bekkers2010framework}.

Artifacts may be structured and stored in different ways depending on the context and process used~\cite{mendezfernandez2012field}. The resulting artifact structure (also called infrastructure) supports communication between different roles and departments inside an organization, e.g., to which product platform a certain feature belongs, what requirements a certain feature consists of, what test cases that belong to a certain requirement, which release a certain requirement should be implemented in, or what artifacts patches that represent the implementation of a certain requirement. The communication schema should be altered dependent on the firms' needs and processes~\cite{fernandez2010meta}, e.g. to follow-up what requirements are contributed. In this study, we introduce an information meta-model that proposes how a set of repositories may be set up to support the above-mentioned communication and decision-making.

Firms often store software artifacts in a central database and require certain quality criteria in terms of completeness and traceability etc~\cite{alspaugh2013ongoing}. In contrast, OSS ecosystems constitute an opposite extreme with their usually very informal practices~\cite{ernst2012case}. Here, a requirement may be represented by several artifacts, often complementing each other to give a more complete picture, e.g., as an issue, in a mail thread, and/or as a prototype or a finished implementation. These artifacts are examples of what Scacchi refers to as informalisms~\cite{scacchi2002understanding} and are stored in decentralized repositories (such as issue trackers, mailing lists, and source code repositories respectively).

\subsection{Summary}
Software engineering has received limited attention in the context of OI, specifically in relation to OSS, which is widespread in practice~\cite{munir2015open}. Hence, the limited attention that contribution strategies have gotten is not surprising with some exceptions~\cite{wnuk2012can, stam2009when}. There is literature explaining general incentives and strategies for how firms should act~\cite{dahlander2008firms, west2006challenges, henkel2014emergence}, but neither of the aforementioned or existing models~\cite{van2009commodification, Bosch13} consider aspects specific to OSS, and how firms should synchronize internal product strategy and planning with OSS ecosystem participation~\cite{munir2015open}. This study aims to address this research gap through a close academia and industry collaboration.

\section{Research methodology}
\label{sec:Design}
In this section, we describe the research design, the process of our study, and our research questions. Further, we motivate the choices of research methods and how these were performed to answer the research questions. Finally, we discuss related validity threats and how these were managed.

\subsection{Case Firm}
\label{subsec:CaseFirm}
Sony Mobile is a multinational firm with roughly 5,000 employees, developing mobile phones and tablets. The studied branch is focused on developing Android based phones and tablets and has 1600 employees, of which 900 are directly involved in software development. Sony Mobile develops software using agile methodologies and uses software product line management with a database of more than 20,000 features suggested or implemented across all product lines~\cite{pohl2005software}.

As reported in earlier work~\cite{muniropen2016}, Sony Mobile is a mature OSS player with involvement in several OSS projects. Their existing processes for managing contribution strategies and compliance issues is centrally managed by an internal group referred to as their OSS governance board~\cite{muniropen2016} (cf. OSS Working group~\cite{kemp2010open}). The board has a cross-functional composition as previously suggested with engineers, business managers, and legal experts, and applies the reactive approach as described in section~\ref{subsec:Reactive}.

\subsection{Research Questions} 
This study aims to support software-intensive firms involved in OSS ecosystems with integrating their internal product strategy and planning~\cite{fricker2012software} with the decision-process of what software artifacts that they should contribute to the OSS ecosystems, and when, formalized as contribution strategies~\cite{wnuk2012can}. Strategic product planning in OI primarily concerns what parts should be revealed (contributed) in an inside-out direction~\cite{chesbrough2003open} from the firm to the ecosystem. This contribution affects the OSS which in turn is sourced in an outside-in direction~\cite{chesbrough2003open} from the ecosystem to the firm and is a key enabler in achieving the potential benefits of OI~\cite{munir2015open}. Earlier research in this area of OI~\cite{west2013leveraging}, and OSS~\cite{munir2015open}, is sparse and often limited to a management level (e.g.,~\cite{dahlander2005relationships, dahlander2008firms, henkel2006selective, van2009commodification}). To occupy this research gap, we aim to design a solution that supports firms in strategic product planning. We pose our first research question (\textbf{RQ1}) as:

\begin{enumerate}[\textbf{RQ1},leftmargin=0.9cm]
\item How can contribution strategies be created and structured to support strategic product planning from an OI perspective?
\end{enumerate}

Product planning is a broad practice and usually involves a cross-functional set of internal stakeholders (e.g., legal, marketing, product management, and developers)~\cite{Komssi2015}. This is also the case for strategic product planning and associated contribution strategies. For a firm with a small development organization, these internal stakeholders may be co-located and efficiently communicate and discuss decisions on a daily basis, but for larger (geographically-distributed) development organizations this may not be possible and cumbersome~\cite{damian2007stakeholders}. A contribution strategy for a certain feature needs to be communicated from the product planning team to the development teams who should implement and contribute accordingly. 
Conversely, product planning is responsible for monitoring the realization of the approved contribution strategies and what impact they have. 

One of the main challenges for market-driven firms is to know what requirements-associated information to obtain, store, manage, and how to enable efficient communication across all stakeholders involved in the crucial decisions that lead to product success~\cite{Karlsson2007588,regnell2005market}. Handling information overload~\cite{Wnuk2011} and efficiently connecting the necessary bits and pieces of information is important for strategy realization and follow up analysis. This is particularly important when introducing new concepts that require close collaboration and efficient communication between product management and product development organizations. Thus, RQ2 focuses on the information meta-model that should be integrated into the software artifact repositories used for requirements management and product planning. Our goal is to develop an information meta-model that describes how contributions to OSS ecosystems can be traced to internal product requirements and platforms, and vice versa, and allow for an organizational adoption of contribution strategies for concerned firms. This leads us to pose our second research question (\textbf{RQ2}):

\begin{enumerate}[\textbf{RQ2},leftmargin=0.9cm]
\item What software and product planning artifact types and repositories are required and how should they be represented in a meta-model to enable communication and follow-up of contribution strategies in strategic product planning?
\end{enumerate}

By answering these two research questions our goal is to create a practical solution for uncovering further benefits that OI brings~\cite{munir2015open}.

\subsection{Research Design and Operation}
\begin{figure}[hbtp]
 \centering
 \includegraphics[width=\textwidth]{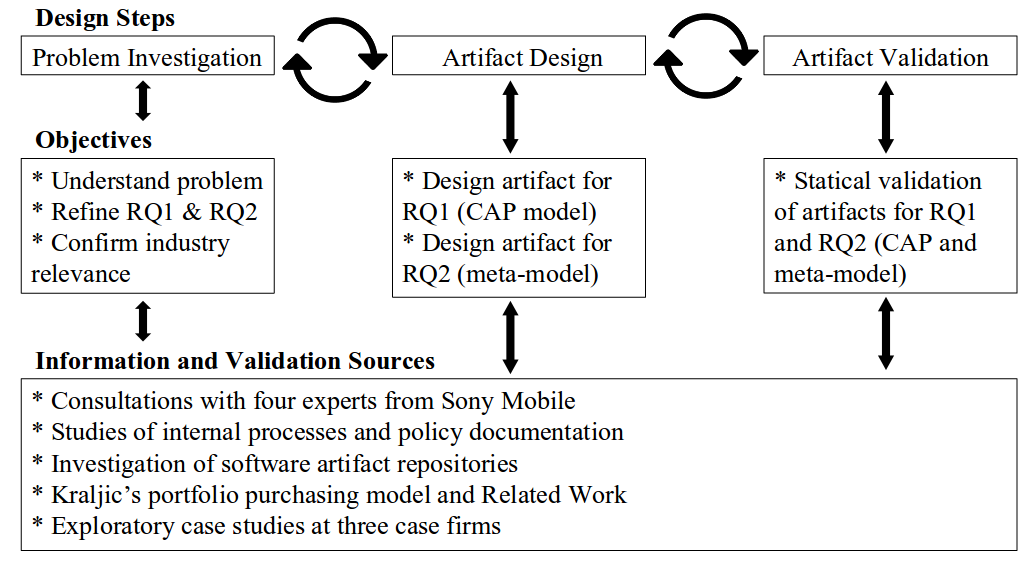}
 \caption{Overview of the research methodology used in this study. The design process was performed iteratively through the three steps involved: problem investigation, artifact design, and artifact valuation~\cite{Hevner2004DSI}.}
 \label{fig:RM}
 \end{figure}

This study is a design science~\cite{Hevner2004DSI} inspired case study~\cite{Runesoncasestudy2012}. The work was initiated by problem identification and analysis of its relevance. This was followed by an artifact design process where the artifacts (the CAP model and information meta-model) addressing the research problems (\textbf{RQ1 \& RQ2}) was created. Finally, the artifacts were validated in the context of the research problem. These steps were performed in close academia-industry collaboration between the researchers and Sony Mobile. We performed data collection and analysis throughout the steps and concluded with reporting of the results (see Fig.~\ref{fig:RM}).


\subsubsection{Problem Identification}
The objectives of the problem investigation phase in the design process~\cite{Hevner2004DSI} are to further understand the problem context and current practices. To gain greater understanding, we conducted informal consultations with four experts (I1-I4) at Sony Mobile who is involved in the decision-making process of OSS contributions (see Table~\ref{tbl:Intervieweeexperts}). This allowed us to further refine both \textbf{RQ1} and \textbf{RQ2} and confirmed their importance and relevance for the industry. Simultaneously, internal processes and policy documentation at Sony Mobile were studied. Next, we received permission to access additional data sources and were able to investigate requirements and contribution repositories. The consultations and investigations confirmed that a suitable solution requires a combination of a technology-based artifact and an organization-based artifact (see guidelines one and two by Hevner~\cite{Hevner2004DSI}). The technology-based artifact (\textbf{RQ1}) should allow firms to create contribution strategies for software artifacts and the organizational-based artifact (\textbf{RQ2}) should support the organizational adoption and operationalization of the technology-based artifact. 

\begin{table}[htbp]
  \centering
  \caption{Consultation with experts}
    \begin{tabular}{lll}
    \toprule
    \textbf{Expert Id } & \textbf{Years of experience} & \textbf{Role}  \\
    \midrule
    I1 & 6 Years & Team Lead  \\
    I2 & 8 Years & Director OSS SW Operations\\
    I3 & 15 Years & Senior Manager  \\
    I4 & 5 Years & Software Developer  \\
    \bottomrule
    \end{tabular}%
  \label{tbl:Intervieweeexperts}%
\end{table}%

\subsubsection{Artifact Design}

\textbf{RQ1} is addressed by designing an artifact that would allow the practitioners to decide whether a software artifact should be contributed to an OSS ecosystem or not. As this is a sourcing issue at the product strategy-level~\cite{kittlaus2008software, fricker2012software,Badampudi2016105}, we decided to base the artifact on Kraljic's portfolio purchasing model~\cite{kraljic1983purchasing} following the advice and experience of I2 in sourcing. The model consists of a matrix that allows firms to analyze how they source and purchase material and components for their production (see section~\ref{subsec:Kraljic}). 

With this foundation, we iteratively formalized our findings from the consultations with I1-I4 and studies of internal processes and policy documentation. The results of this formalization are the CAP model and the associated meta-model of information required to instantiate the CAP model, supporting strategic product planning in OI. Each item category from the original model~\cite{kraljic1983purchasing} has a corresponding type of contribution strategy~\cite{wnuk2012can}, and instead of supply items, we refer to software artifacts, e.g., features or components. The two dimensions are refined to represent \textit{Business impact} and \textit{Control complexity}, inspired by existing commoditization models~\cite{van2009commodification, Bosch13} and literature on OSS ecosystem governance (e.g.,~\cite{baars2012framework, dahlander2005relationships,nakakoji2002evolution}). The measurement process is proposed to employ a consensus-seeking approach \cite{gelderman2003handling} with the involvement of cross-functional competencies and internal stakeholders~\cite{Komssi2015}. To help frame the measurement discussion process, questions are defined inspired by literature related to the Kraljic portfolio purchasing model (e.g.,~\cite{gelderman2003handling,Caniels2005141}), commoditization models~\cite{van2009commodification, Bosch13}, software value map (e.g.,~\cite{khurum2013software, aurum2007value}, and OSS ecosystem governance (e.g.,~\cite{baars2012framework, dahlander2005relationships,nakakoji2002evolution}). An overlay is created on top of the CAP model to highlight which contribution objective should be the primary driver for the chosen contribution strategy. The objectives represent important value incentives inspired by OI literature~\cite{chesbrough2003open, west2003open, stuermer2009extending, ven2008challenges}. The intention is to help users of the model to fine-tune the contribution strategy for the classified artifact. The CAP model is presented in more detail in section~\ref{sec:Model}.

To address \textbf{RQ2} and enable an organizational adoption and operationalization of the CAP-model, we created an information meta-model that facilitates communication and follow-up on software artifacts and their contribution strategies. In the problem investigation phase, it became apparent that the information support should be integrated into the software artifact repositories used for requirements management. The information support would then be able to reach everyone who is involved in the product planning and development. This required us to expand our investigation of Sony Mobile's requirements and contribution repositories, which included a broad set of software artifact repositories that are used in the product planning of mobile phones. We focused the repository investigation on understanding how contributions could be traced to product requirements and platforms, and vice versa. Through consultation with I1-I4, we selected six relevant repositories: the internal product portfolio, feature repository, feature-based architectural asset repository, patch repository, contribution repository and commit repository (see section~\ref{sec:Repositories}). 

These repositories and their unique artifact IDs (e.g., requirement id, patch id, and contribution id) allowed us to trace the contributions and commits to the architectural assets, product requirements and platforms, via the patches that developers create and commit to internal source code branches. This analysis resulted in the information meta-model presented in Fig.~\ref{fig:SoftwareRepositories}. The meta-model creation process was driven by the principles of finding a balance between research rigor and relevance, moving away from extensive mathematical formalizations of the CAP model and focusing on the applicability and generalizability of the model, see guideline five by Hevner~\cite{Hevner2004DSI}.

\subsubsection{Artifacts Validation}
Validation helps confirm that candidate solutions actually address the identified research problems. As we are in an early stage of the research and design process, this study uses static validation~\cite{gorschek2006model}. This type of validation uses presentation of candidate solutions to industry practitioners and gathering of feedback that can help to further understand the problem context and refine candidate solutions, in line with the design science process~\cite{Hevner2004DSI}. Dynamic validation~\cite{gorschek2006model}, which concerns piloting of the candidate solutions in a real-work setting, is a later step in the technology transfer process and is currently under planning at the case firm and is left for future work. 

Both the CAP-model and its related information meta-model were validated statically through continuous consultations with experts at Sony Mobile (I1-I4). In these consultations, the models were explained and discussed. Feedback and improvement ideas were collected and used for iterative refinement and improvement. Experts were asked to run the CAP model against examples of features in relation to the four software artifact categories and related contribution strategies that CAP model describes. The examples are presented together with the CAP model and provide further detail and validation of its potential use, see section~\ref{sec:contributionStrategies}. A complete example of how the CAP model and meta-model are used is further presented in section~\ref{sec:Example}. These examples help to evaluate functionality, completeness, and consistency of the CAP model and associated information meta-model. The usability of the information meta-model was further validated by performing traces between the different types of artifacts and their repositories. These traces were presented and used in the static validation of the meta-model. From a design science perspective~\cite{Hevner2004DSI}, we employed observational validation through a case study at Sony Mobile where we studied the artifacts (models) in a business environment. We also employed descriptive evaluation where we obtained detailed scenarios to demonstrate the utility of the CAP model, see guideline three by Hevner~\cite{Hevner2004DSI}.

To improve the external validity of the CAP model, we conducted exploratory case studies at three different case firms (see Section~\ref{sec:CaseStudies}). In these case studies, we used static validation~\cite{gorschek2006model} where we presented the CAP model to participants from the respective firms and applied it in a simulated setting as part of the interviews. In two of the cases, semi-structured interviews were used with one representative from each firm. In the third case, a workshop setting was used with eight participants from the firm. When collecting feedback from the three case firms, we focused on applicability and usability of the CAP model.

\subsection{Ethics and Confidentiality}
\label{sec:ethics}
This study involved analysis of sensitive data from Sony Mobile. The researchers in the study had to maintain the data's integrity and adhere to agreed procedures that data will not be made public. Researchers arranged meetings with experts from Sony Mobile to inform them about the study reporting policies. Data acquired from Sony Mobile is confidential and will not be  publicly shared to ensure that the study does not hurt the reputation or business of Sony Mobile. Finally, before submitting the paper for publication, the study was shared with an expert at Sony Mobile who reviewed the manuscript to ensure the validity and transparency of results for the scientific community.

\subsection{Validity Threats}
\label{sec:Validity Threats}
This section highlights the validity threats associated with the study. Four types of validity threats~\cite{Runesoncasestudy2012} are mentioned along with their mitigation strategies.

\subsubsection{Internal Validity}
\label{sec:Internalvalidity}
Internal validity refers to factors affecting the outcome of the study without the knowledge of the researchers~\cite{Runesoncasestudy2012}. 

\textit{Researcher bias} refers to when the researcher may risk influencing the results in a wanted direction~\cite{runeson2009guidelines}. The proposed CAP model was created with an iterative cooperation between researchers and industry practitioners. Thus, there was a risk of introducing the researcher's bias while working towards the creation of the model. In order to minimize this risk, regular meetings were arranged between researchers and industry experts to ensure the objective understanding and proposed outcomes of the study. Furthermore, researchers and industry practitioners reviewed the paper independently to avoid introducing researcher's bias. 

A central part of the CAP model involves estimating the business impact and control complexity. These estimations involve several factors and can have multiple confounding factors that influence them. In this work, we assume that this threat to internal validity is taken into consideration during the estimation process and therefore is not in the direct focus of the CAP model. Moreover, the CAP model does not prevent additions of new factors that support these estimates. 

\textit{Triangulation} refers to the use of data from multiple sources and also ensuring observer triangulation~\cite{runeson2009guidelines}. In this study, our data analysis involved interpretation of qualitative and quantitative data obtained from Sony Mobile. We applied data triangulation by using Sony Mobile's internal artifacts repositories, documents related to contribution strategies and consultation with relevant experts before proposing the CAP model. There were risks of identifying the wrong data flows and subjective interpretation of interviews. In order to mitigate these risks, concerned multiple experts with different roles and experiences (see Table~\ref{tbl:Intervieweeexperts}) were consulted at Sony Mobile. We ensured observer triangulation by involving all researchers who authored this manuscript into the data collection and analysis phases. 

\subsubsection{External Validity}
\label{sec:Externalvalidity}
External validity deals with the ability to generalize the study findings to other contexts. 

We have focused on analytic generalization rather than statistical generalization~\cite{Flyvbjerg07} by comparing the characteristics of the case to a possible target and presenting case firm characteristics as much as confidentiality concerns allowed. The scope of this study is limited to firms realizing OI with OSS ecosystems. Sony Mobile represents an organization with a focus on software development for embedded devices. However, the practices that are reported and proposed in the study has the potential to be generalized to all firms involved in OSS ecosystems. It should be noted that the case firm can be considered a mature firm in relation to OSS usage for creating product value and realizing product strategies. Also, they recognize the need to invest resources in the ecosystems by contributing back in order to be able to influence and control in accordance with internal needs and incentives. Thus, the application of the proposed CAP model in an other context or in other firms remains part of future work.  

The CAP model assumes that firms realize their products based, in part, on OSS code and OSS ecosystem participation. This limits its external generalizability to these firms. At the same time, we believe that the innovation assessment part of the CAP model may be applied to artifacts without OSS elements. In this case, the CAP model provides only partial support as it only helps to estimate the innovativeness of the features (as an innovation benchmark) without setting contribution strategies. Still, this part of the CAP model should work in the same way for both OSS and non-OSS based products. Finally, the classification of software artifacts has a marked business view and a clear business connotation. A threat remains here that important technical aspects (e.g. technical debt, architectural complexity) are overlooked. However, throughout the static validation examples, we saw limited negative impact on this aspect, especially in a firm experienced in building its product on an OSS platform.  

The meta-model was derived from Sony Mobile's software artifact repositories. We believe that the meta-model will fit organizations in similar characteristics. For other cases, we believe that the meta-model can provide inspiration and guidance for how development organizations should implementing the necessary adaptations to existing requirements management infrastructure, or create such, so that contribution strategies for artifacts can be communicated and monitored. We do acknowledge this as a limitation in regards to external validity that we aim to address in future design cycles.

\subsubsection{Construct Validity}
\label{sec:ConstructValidity}
Construct validity deals with choosing the suitable measures for the concepts under study~\cite{runeson2012casestudy}. Four threats to the construct validity of the study are highlighted below. 

First, there was a risk that academic researchers and industry practitioners may use different terms and have different theoretical frames of reference when addressing contribution strategies. Furthermore, the presence of researchers may have biased the experts from Sony Mobile to give information according to researchers' expectations. The selection of a smaller number of experts from Sony Mobile might also contribute to the unbalanced view of the construct.

Second, there was a potential threat to construct validity due to the used innovation assessment criteria based on business impact and control complexity. Both dimensions can be expanded by additional questions (e.g. internal business perspective or innovation and learning perspective~\cite{SMR1560}) and the CAP model provides this flexibility. One could argue that also technical and architectural aspects should be taken into consideration here. At the same time, the static validation results at Sony Mobile confirm that these aspects have limited importance at least for the studied cases. Still, they should not be overlooked when executing the CAP model in other contexts.  

Third, a common theoretical frame of reference is important to avoid misinterpretations between researchers and practitioners~\cite{runeson2009guidelines}. In this study, the Kraljic's portfolio model is used as a reference framework to the CAP model. However, the horizontal and vertical dimensions of Kraljic's portfolio model were changed to control complexity and business impact respectively. Both industry practitioners and academic researchers had a common understanding of Kraljic's portfolio model~\cite{kraljic1983purchasing} before discussions in the study. Furthermore, theoretical constructs were validated by involving one of the experts in the writing process from Sony Mobile to ensure consistent understanding.

Fourth, prolonged involvement refers to a long-term relationship or involvement between the researchers and organization~\cite{runeson2009guidelines}. Since there was an involvement of confidential information in the study, it was important to have a mutual trust between academic researchers and practitioners to be able to constructively present the findings. The adequate level of trust was gained as a result of long past history of collaboration between academic researchers and experts from Sony Mobile.



\subsubsection{Reliability}
\label{sec:Reliability}
The reliability deals with to what extent the data and the analysis are dependent on the specific researcher and the ability to replicate the study. 

\textit{Member checking} may involve having multiple individuals go through the data, or letting interviewees review a transcript~\cite{runeson2009guidelines}. In this study, the first two authors proposed the meta-model after independent discussions and reviewed by the third author. Furthermore, the model was validated by a team lead, software developer, and senior manager at Sony Mobile, involved in making contributions to OSS communities, were consulted to ensure the correctness of the meta-model and associated data. 

\textit{Audit trail} regards maintaining traceability between collected data during the study~\cite{runeson2009guidelines}. For this study, the first two researchers kept track of all the mined data from the software artifact repositories as well as the email and informal communication between researchers and Sony Mobile representative. Results were shared with Sony Mobile for any possible misinterpretation or correction of data.

\section{The Contribution Acceptance Process (CAP) Model (RQ1)}
\label{sec:Model}

The CAP model is an adapted version of the portfolio model introduced by Peter Kraljic~\cite{kraljic1983purchasing}. Kraljic's model was originally constructed to help firms with creating purchasing strategies towards their suppliers of items needed for their product manufacturing. The CAP model is focused on software artifacts and how these should be sourced and contributed as OSS. The artifacts may be of different abstraction levels, e.g., ranging from specific requirements or issues to sets of requirements as features, frameworks, tools or higher level components.

The model may be used \textit{proactively} or \textit{reactively}. In the former, the model is systematically used on a portfolio or set of artifacts to decide on specific contribution strategies for each artifact, but also to get a general overview and analyze the artifacts relative each other. In the reactive case, the model is used to follow-up on previously classified artifacts, and for individual contribution requests of artifacts from the development organization. We start by describing how the model may be used to classify artifacts and elicit contribution strategies. We then move on and put the model into the context of the two approaches. Lastly, we give examples of artifacts and related contribution strategies.

\subsection{Model Description}
The focal point of the CAP model is the matrix presented in Fig.~\ref{fig:OImodel}. Artifacts are mapped on to the matrix based on how they are valued in regard to the two dimensions \textit{Business impact} and \textit{Control complexity}, located on the vertical and horizontal axis respectively. Business impact refers to how much you profit from the artifact, and control complexity refers to how hard the technology and knowledge behind the artifact is to acquire and control. Both dimensions range from low to high.  

\subsubsection{Artifact Types and Contribution Strategies}
An artifact is categorized into one of the four quadrants, where each quadrant represents a specific artifact type with certain characteristics and contribution strategy. The four types are as follows:

\begin{figure*}
\begin{center}
\includegraphics[width=\textwidth]{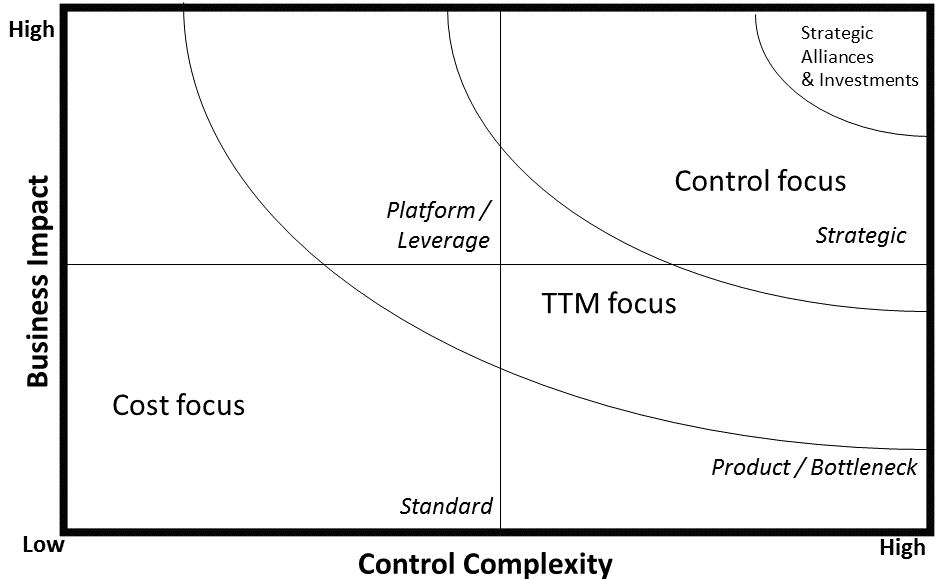}
\label{fig:OImodel}
\caption{The Contribution Acceptance Process (CAP) model and its different quadrants that help to determine what contribution strategy to use depending on how a software artifacts are classified in terms of \textit{business impact} and \textit{control complexity}. The overlaying arches marks up four contribution objectives which help to further tailor the contribution strategy (see section~\ref{sec:ContributionObjectives}).}
\end{center}
\end{figure*}

\begin{itemize}
    \item Strategic artifacts: high business impact and high control complexity.
    \item Platform$/$leverage artifacts: high business impact and low control complexity.
    \item Products$/$bottlenecks artifacts: low business impact and high control complexity.
    \item Standard artifacts: low business impact and low control complexity.
\end{itemize} 

\textbf{Strategic Artifacts:} This category includes artifacts that can be internally or externally developed, have a differential value and makes up a competitive edge for the firm. Due to their value and uniqueness, there is a need to maintain a high degree of control over these artifacts. OSS contributions within this category should generally be restricted and made in a controlled manner, ensuring that the differentiation is kept. However, this does not account for possible enablers and/or frameworks, i.e., parts of the artifact that are required for the artifact to work in a given environment. Those have to be actively maintained and contributed. This may require that the artifacts undergo special screening to identify the parts that enable the differentiating parts.
In case the artifact is already connected to an existing OSS ecosystem, the firm should strive towards gaining and maintaining a high influence in the ecosystem in regard to the specific artifact and attached functionality. If this is not achievable, e.g., when the contribution terms of an existing ecosystem require contributions to include the differential IP, the option of creating a new and firm-orchestrated OSS ecosystems should be considered. For examples of Strategic artifacts, see section~\ref{subsec:Strategic Artifacts}.

\textbf{Platform/Leverage Artifacts:} These artifacts have a high degree of innovation and positive business impact, but their development does not necessarily need to be controlled by the firm. Examples include technology and market opportunity enablers that have competing alternatives available, ideally with a low switching cost. Generally, everything could be contributed, but with priority given to contributions with the highest potential to reduce time-to-market, i.e., contributions with substance should be prioritized over minor ones, such as error-corrections and maintenance contributions that are purely motivated due to cost reduction. Due to the lower need for control, firms should strive to contribute to existing projects rather than creating new ones, which would require a substantial degree of effort and resources and represent an unnecessary investment. For examples of Platform/Leverage artifacts, see section~\ref{subsec:Platform/Leverage Artifacts}.

\textbf{Products/Bottleneck Artifacts:} This category includes artifacts that do not have a high positive business impact by itself but would have a negative effect if not present or provided. For example, functionality firmly required in certain customer-specific solutions but are not made available for the general market. These artifacts are hard to acquire and requires a high degree of control due to the specific requirements. The strategy calls for securing the delivery for the specific customers, while and if possible, sharing the burden of development and maintenance. Generally, everything could be contributed, but with priority given to contributions with the highest potential to reduce time-to-market, or in this case rather the time-to-customer. But, due to the unique nature of these artifacts, the number of other stakeholders may be limited in existing OSS ecosystems. This may imply that the artifact will be problematic to contribute in a general OSS ecosystem. An option would then be to identify and target specific stakeholders of interest, i.e. of customers and their suppliers, and create a limited project and related OSS ecosystem. For examples of Products/Bottlenecks artifacts, see section~\ref{subsec:Bottleneck Artifacts}.

\textbf{Standard Artifacts:} This category includes artifacts that may be considered as a commodity to the firm. They do not have a competitive edge if kept internal and has reached a stage in the technology life-cycle where they can create more value externally. They may be externally acquired as easily as internally developed and may, therefore, be considered to have a low level of control complexity. Generally, everything should be contributed, but with priority given to contributions with the highest cost reduction potential. Creating a competing solution to existing ones could lead to unnecessary internal maintenance costs, which has no potential of triggering a positive business impact for a firm. For examples of Standard artifacts, see section~\ref{subsec:Standard Artifacts}.

\subsubsection{Contribution Objectives}
\label{sec:ContributionObjectives}
Mapping an artifact relative to the four quadrants brings an indication and guideline about its contribution strategy. There are also intrinsic objectives for making contributions that are not fully captured by just accessing the business impact and control complexity in the artifact classification process. These objectives include:

\begin{itemize}
    \item Cost focus
    \item Time-to-market (TTM) focus
    \item Control focus
    \item Strategic Alliances and Investments
\end{itemize}

These objectives are closely coupled to the different strategies and are presented as an overlay of the matrix, thus emphasizing the main contribution objective per strategy.

\textbf{Cost focus:} Artifacts with a limited competitive advantage, i.e., they are considered as commodity or enablers for other artifacts, will have a contribution objective mainly focused on \textit{reducing the cost of development and maintenance}. The contribution strategy should focus on minimizing the number of internal patches that need to be applied to each new OSS project release and reusing common solutions available in OSS to fulfill internal requirements, i.e., overall reduce variants and strive for the \textit{standardization} that comes with OSS. As a consequence, internal resources may be shifted towards tasks that have more differentiation value for a firm.  

\textbf{Time-To-Market (TTM) focus:} Artifacts that have higher levels of competitive advantage, and/or require a higher amount of control and understanding than commodity artifacts should likely to have the general objective to be advanced to the marketplace as soon as possible, superseding the objective of reducing maintenance costs. These artifacts may also be referred to as \textit{qualifiers}, i.e., artifacts that are essential but still non-differential, and should be contributed as soon and often as possible in order to allow for the own solution to be established as the leading open solution. This will potentially give the advantage of control and barring competing solutions which would otherwise require additional patching or even costly redesigns to one's own product. 

\textbf{Control focus:} Artifacts with a high level of competitive advantage and requiring a high level of control are likely to provide differentiation in the marketplace, and should thus not be contributed. Yet, in securing that these artifacts are enabled to operate in an open environment, it is as important to contribute the enabling parts to the OSS ecosystems. If an alternative open solution would become widely adapted out of the firm's control, the firm's competitive edge will likely be diminished and make a costly redesign imperative. Hence, the contribution objective for these artifacts is to take control of the OSS ecosystem with the general strategy to gain and maintain necessary influence in order to better manage conflicting agendas and levy one's own strategy in supporting the artifact.

\textbf{Strategic Alliances and Investments:} These artifacts carry a very large part of product innovation and competitive advantage, and require strict control. Thus, these artifacts should be internally developed, or, if this is not feasible, co-developed using strategic alliances and investments that secure IPR ownership, hence there is generally no objective for making open source contributions.

\subsubsection{Adapting Contribution Strategies with Contribution Objectives}
Having just a single contribution objective for an artifact is rare except for the extreme cases, e.g., when an artifact is mapped in the far corners of the matrix, such as the bottom left as strictly standard and commodity. More common is to have two or more contribution objectives in play, though one of the objectives would be the leading one. The overlay of contribution objectives on the matrix's different contribution strategies is intended as a guidance for fine-tuning the contribution strategy for individual artifacts when more than one contribution objective is in play. E.g., although two artifacts who are found to have the same overall Platform/Leverage contribution strategy, there might be a degree of difference in the emphasis to be made in the time-to-market objective for an artifact closer to the Strategic area, compared with an artifact closer to the Standard area where considerations on cost of maintenance might overtake as the leading objective.

\subsection{Proactive Approach}
\label{subsec:Proactive}
When proactively using the model, the following step-by-step approach is recommended:

\begin{enumerate}[label=S\arabic*]
    \item Decision on scope and abstraction level.
    \item Classification and mapping artifacts to the matrix.
    \begin{enumerate}
        \item Begin with an initial set of artifacts to the matrix.
        \item Synchronize and reiterate mapping.
        \item Map the rest of the artifacts to the matrix.
    \end{enumerate}
    \item Reiteration of the artifact mapping.
    \item Documentation and communication of the decisions.
    \item Monitoring and follow-up on the decisions.
\end{enumerate}

Before the model is used, the scope and abstraction level of the analysis needs to be decided (\textbf{S1}). The scope may, for example, entails a product, a platform or functional area. Abstraction level concerns the artifacts relative to the scope, e.g., components, features, or requirements. Based on these limitations, the artifacts should be listed, and necessary background information collected, e.g., market intelligence, architectural notes and impact analysis, OSS ecosystem intelligence, and license compliance analysis. 

The collected information should then be used as input to an open consensus-seeking discussion forum (\textbf{S2}), where relevant internal stakeholders help to classify the artifacts. As in the roadmapping process~\cite{Komssi2015}, these stakeholders should bring cross-functional perspective to the decision-making to further explain and argue based on the collected background information, e.g., representatives from marketing, product management, development, and legal.

To facilitate the discussions and help assess the business impact of the artifacts, a set of questions may be used. The joint answers to these questions are given on a Likert scale with values between 1 and 4. The reason for this scale is to force discussion participants to take a clear stand on which side of two quadrants they think an artifact belongs. The questions are as follows (it equals an artifact):

\begin{enumerate}
    \item How does it impact on the firm's profit and revenue?
    \item How does it impact on the customer and end user value?
    \item How does it impact on the product differentiation?
    \item How does it impact on the access to leading technology/trends?
    \item How does it impact if there are difficulties or shortages?
\end{enumerate}

As with the business impact, a set of questions are proposed to help asses the control complexity of the artifact on a scale between 1-4:

\begin{enumerate}
    \item Do we have knowledge and capacity to absorb the technology?
    \item Are there technology availability barriers and IPR constraints?
    \item What is the level of innovativeness and novelty?
    \item Is there a lack of alternatives?
    \item Are there limitations or constraints by the firm?
\end{enumerate}

For an example of how these questions can be used, see section~\ref{sec:Example}. When all questions are answered, the mean values for both dimensions should be calculated. Based on these values, the artifact is then mapped onto the matrix (see Fig.~\ref{fig:OImodel}), which will put it into one of the four quadrants. The group should then ask themselves if the calculated position agrees with their general belief of where it should be. They should also ask themselves where they want it to be. Further, they should consider what contribution objective(s) that apply, and how this affects the contribution strategy. This process should be allowed to take time and reiteration of the first set of artifacts, as this is necessary for everyone to get accustomed with the process and the classification criteria.

This classification process is not intended to be quantitative and rigorous, but rather qualitative and informal. The process was facilitated through consensus-seeking discussions within a cross-functional group. This approach helps to create guidelines without introducing complexity which may risk introducing negative effects on the usability and applicability of the CAP model. The questions should further be seen as a mean to frame and drive the discussion, during which further questions might come up.

When all artifacts have been classified and mapped onto the matrix, an overall discussion and reflection should be performed (\textbf{S3}). When consensus is reached, the decisions should be documented and handed over to product management for communication out to the development organization (\textbf{S4}) through required channels supported by the information meta-model, e.g., the requirements management infrastructure (see section~\ref{sec:Repositories}). The contribution strategies for each artifact should then be monitored and followed-up in a given suitable time frame (e.g., in relation to internal release cycles) (\textbf{S5}). This task may be suitable for product or project management with accountability towards the firm's OSS executive.

\subsection{Reactive Approach}
\label{subsec:Reactive}
The CAP model may also be used in a \textit{reactive} mode which is based on Sony Mobile's current practices. This approach is critical in order to continuously follow-up on previously classified artifacts as the classification may change with the artifacts' technology life-cycle. The approach is also useful for managing individual contribution requests of artifacts from the development organization, e.g. in response when a manager or developer request to contribute a certain artifact, or be allowed to work actively with a specific OSS ecosystem. The CAP model is used in this case by a group of internal stakeholders, similarly to that of the proactive approach. Sony Mobile applies this reactive approach through their OSS governance board (see section~\ref{subsec:CaseFirm}).

When an individual wants to make a contribution, they have to pass through the board. However, to avoid too much bureaucracy and a bottleneck effect, the contribution process varies depending on the size and complexity of the contribution. In the CAP model, the contributions may be characterized in one of three different levels:

\begin{itemize}
    \item \textbf{Trivial contributions} are rather small changes to already existing OSS ecosystems, which enhances the non-significant code quality without adding any new functionality to the system e.g., bug fixes, re-factoring etc. 
    \item \textbf{Medium contributions} entails both substantially changed functionality, and completely new functionality e.g., new features, architectural changes etc.
    \item \textbf{Major contributions} are comprised of substantial amounts of code, with significant value in regard to IPR. These contributions are a result of a significant amount of internal development efforts. At Sony Mobile, one example of such a contribution is the Jenkins-Gerrit-trigger plug-in~\cite{muniropen2016}.
\end{itemize}

For trivial contributions, the approval of concerned business manager is sufficient. For medium and major contributions, the business manager has to prepare a case for the Open Source Governance board to verify the legal and IPR aspects of the OSS adoption or contribution. The Open Source Governance board decides after case investigation that include IPR review. Consequently, the board accepts or rejects the original request from the engineers. To further lessen the bureaucracy, Sony Mobile uses frame agreements that can be created for OSS ecosystems that are generally considered as having a non-competitive advantage for Sony Mobile (e.g., development and deployment infrastructure). In these cases, developers are given free hands to contribute what they consider as minor or medium contributions, while major contributions must still go through the board.

\subsection{Contribution Strategies with Artifact Examples}
\label{sec:contributionStrategies}
In this section, we provide examples in regard to the four artifact types of the CAP model, which we elicited from consultations with experts from Sony Mobile.

\subsubsection{Strategic Artifacts:} 
\label{subsec:Strategic Artifacts}
\textbf{Example 1 - Gaming, Audio, Video, and Camera:}
A typical example of a contributable enabler is multimedia frameworks which are needed for services such as music, gaming, and videos. The frameworks themselves are not of a strategic value, but they are essential for driving the Sony brand proposition since they are needed in order to provide the full experience of strategic media and content services provided by Sony. Such artifacts may also be referred to as Qualifiers, as they are essential, yet not strategic by themselves.

An example of such a multimedia framework that Sony Mobile uses is Android’s Stagefright\footnote{https://source.android.com/devices/media/}. It is for example used for managing movies captured by the camera. The framework itself could be contributed into, but not specific camera features such as smile recognition as these are considered as differentiating towards the competition, hence have a high business impact and control complexity for Sony Mobile. In short, camera effects can not be contributed, but all enablers of such effects should be, thus Sony Mobile contributes to the frameworks to steer and open up a platform for strategic assets, e.g., an extended camera experience on their mobile phones. A further example of a framework that has been made open by Sony, but in the context of gaming, is the Authoring Tools Framework\footnote{https://github.com/SonyWWS/ATF} for the PlayStation 4.

\subsubsection{Platform/Leverage Artifacts}
\label{subsec:Platform/Leverage Artifacts}
\textbf{Example 1 - Digital Living Network Alliance:} Digital Living Network Alliance (DLNA) (originally named Digital Home Working Group) was founded by a group of consumer electronics firms, with Sony and Intel in leading roles, in June 2003. DLNA promotes a set of interoperability guidelines for sharing and streaming digital media among multimedia devices.

As support for DNLA was eventually included in Android, creating a proprietary in-house solution would not have been wise given that the OSS solution already was offered. Instead, Sony Mobile chose to support the Android DNLA solution with targeted but limited contributions. This is a typical example of leveraging functionality that a firm does not create, own, or control, but that is good to have. Hence, Sony Mobile did not need to commit extra resources to secure the interoperability of an own solution. Instead, those extra resources could be used for making the overall offering better, e.g., the seamless streaming of media between Android devices and other DNLA compliant device, for instance, a PlayStation console, and in that way promote DNLA across Sony’s all device offerings. 

\textbf{Example 2 - Mozilla Firefox:} The most significant web browsers during the 1990s were proprietary products. For instance, Netscape was only free for individuals, business users had to pay for the license. In 1995, Microsoft stepped into browser market due to the competitive threat from Netscape browser. Microsoft decided to drive the price of web browsers market by bundling its competitive browsers for free with the Windows operating system. In order to save the market share, Netscape open sourced the code to its web browsers in 1998 which resulted in the creation of the Mozilla organization. The current browser known as Firefox is the main offspring from that time. By making their browsers open source, Netscape was able to compete against Microsoft's web browsers by commoditizing the platform and enabling for other services and products.

\subsubsection{Products/Bottleneck Artifacts} 
\label{subsec:Bottleneck Artifacts}
\textbf{Example 1 - Symbian network operators requirements:}
In the ecosystem surrounding the Symbian operating system, network operators were considered one of the key stakeholders. Network operators ran the telephone networks to which Symbian smart-phones would be connected. Handset manufactures are dependent on the operators for distribution of more than 90\% of the mobile phone handsets, and they were highly fragmented, with over 500 networks in 200 countries. Consequently, operators can impose requirements upon handset manufactures in key areas such as pre-loaded software and security. These requirements can carry the potential to one of those components that do not contribute in terms of a business value but would make a negative impact on firm's business if missing, e.g., by a product not being ranged.

\textbf{Example 2 - DoCoMo mobile phone operator:}
DoCoMo, an operator on the Japanese market, had the requirement that the DRM protection in their provided handsets uses Microsoft's PlayReady DRM mechanism. This requirement applied to all handset manufacturers, including Sony Mobile's competitors. Sony Mobile, who had an internally developed PlayReady plug-in, proposed that they could contribute it as OSS and create an ecosystem around it and also because it already contributed the DRM framework. DoCoMo accepted, which allowed Sony Mobile and its competitors to share maintenance and development of upcoming requirements from DoCoMo. In summary, Sony Mobile solved a potential bottleneck requirement which has no business value for them by making it OSS and shared the development cost with all its competitors while still satisfying the operator. 

\subsubsection{Standard Artifacts} 
\label{subsec:Standard Artifacts}
\textbf{Example 1 - WiFi-connect\footnote{https://github.com/resin-io/resin-wifi-connect}:}
This OSS checks whether a device is connected to a Wi-Fi. If not, it tries to join the favorite network, and if this fails, it opens an Access Point to which you can connect using a laptop or mobile phone and input new Wi-Fi credentials.

\textbf{Example 2 - Universal Image Loader\footnote{https://github.com/nostra13/Android-Universal-Image-Loader}:} Universal Image Loader is built to provide a flexible, powerful and highly customizable instrument for image loading, caching and displaying. It provides a lot of configuration options and good control over the image loading and caching process.

Both examples are considered standard artifacts because they can be considered as a commodity, accessible for competition and do not add any value to customers in the sense that they would not be willing to pay extra for them.

\section{Operationalization of the CAP model (RQ2)}
\label{sec:Repositories}
Putting contribution strategies into practice requires appropriate processes and information support to know which artifacts, or what parts of them that should be contributed. Furthermore, to follow up the contribution strategy execution and make necessary adaptations as the market changes, there needs to be a possibility to see what has been contributed, where, and when. In this section, we address research question \textbf{RQ2} and propose an information meta-model which can be used to record and communicate the operationalization of the CAP model, e.g., by integrating it into the requirements management and product management information infrastructure. 

The meta-model was created through an investigation of Sony Mobile's software and product management artifact repositories used in product planning and product development. During this investigation, we focused on how the contributions could be traced to product requirements and platforms, and vice versa. Through consultation with I1-4, the investigation resulted in the selection of six repositories, see Fig.~\ref{fig:SoftwareRepositories}:

\begin{figure}[hbtp]
\centering
\includegraphics[width=\textwidth]{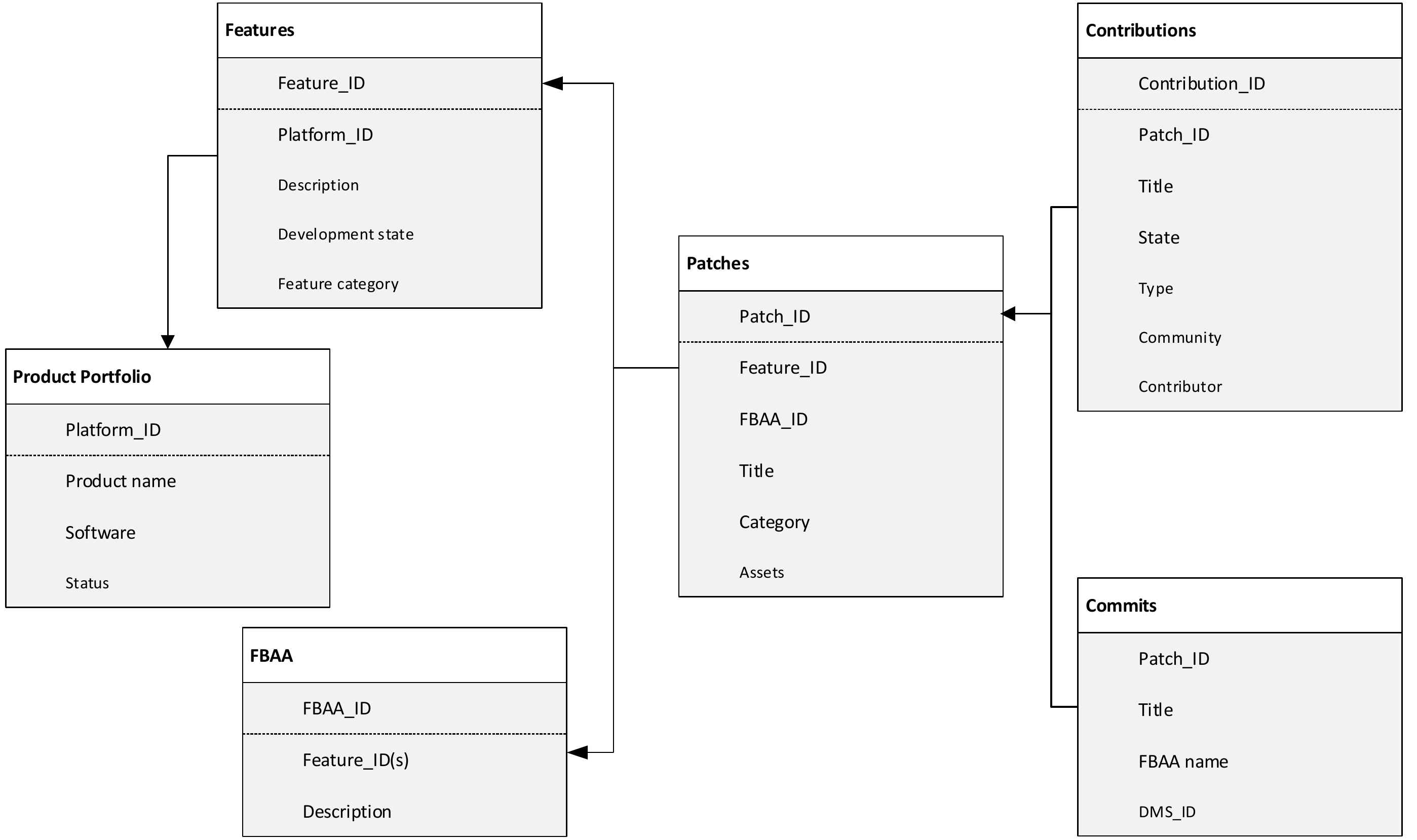}
\caption{Software artifact repositories necessary to communicate and follow-up on contribution strategies decided with the CAP model.}
\label{fig:SoftwareRepositories}
\end{figure}

\begin{itemize}
    \item Product Portfolio repository
    \item Features repository
    \item Feature-Based Architecture Assets repository
    \item Patch repository
    \item Contribution repository
    \item Commit repository
\end{itemize}

\begin{table*}[htbp]\footnotesize
  \centering
  \caption{Description of selected attributes from the software artifact repositories mentioned in Fig.~\ref{fig:SoftwareRepositories}}
  \resizebox{\columnwidth}{!}{
    \begin{tabular}{p{1.7cm} p{2.1cm} p{8cm}}
    \toprule
    \textbf{Repository Name} & \textbf{Attributes} & \textbf{Description} \\
    \midrule
    \multirow{3}[8]{*}{\textbf{Products}} & Platform ID & A unique ID for platform name \\
          & Product name & Product name with the platform. \\
          & Software & Related software description, e.g., Android, OSE, Epice, Kept etc. \\
          & Status & Current standing of the platform, e.g., expired, announced etc. \\
          \midrule
    \multirow{5}[10]{*}{\textbf{Features}} & Feature ID & A unique Id for a feature, which refers to features. \\
          & Platform ID & ID associated with the specific platform e.g. android, core etc. \\
          & Description & Details of the feature. \\
          & Development state & Refers to the current status a feature's implementation, e.g., started, executed. \\
          & Feature category & Refers to the type of feature, e.g., new functionality, bug fix, extension etc. \\
          & Contribution Strategy & Refers to whether the requirement is contributable or not. \\
          \midrule
    \multirow{2}[6]{*}{\textbf{FBAA}} & FBAA ID & A unique Id for each Feature Based Architecture Asset (FBAA). \\
          & FP IDs & A combination of FP IDs associated with the FBAA. \\
          & Description & Details of a FBAA. \\
          \midrule
    \multirow{5}[12]{*}{\textbf{Patches}} & Patch ID & A unique id for each patch. \\
          & FP ID & A unique ID from the FP repository. \\
          & FBAA ID & A unique ID from the FBAA repository. \\
          & Title & A description of a patch. \\
          & Category & Importance of a patch, e.g., market critical, development critical, stability, ecosystem critical etc. \\
          & Assets & Refers to the type of a patch, e.g., bug fix, extension, operator requirement, platform related, generic etc. \\
          \midrule
    \multirow{7}[14]{*}{\textbf{Contributions}} & Contribution ID & A unique ID for each contribution. \\
          & Patch ID & A unique ID from the patches repositories. \\
          & Title & A description of a contribution. \\
          & State & Refers the current state of the patch, e.g., ecosystem merged, already fixed, CEO rejected, legal reject, ecosystem review etc. \\
          & Type  & Refers to criticality of a contribution, e.g., trivial, non-trivial, bug fix etc.  \\
          & ecosystem & Refers to the ecosystem in which the contribution will be made, e.g., Google, Firefox etc. \\
          & Contributors & Refers the contributor information. \\
          \midrule
    \multirow{1}[8]{*}{\textbf{Commits}} & Patch ID & A unique Id from the patch repository. \\
          & Title & A detailed description of a commit. \\
          & FBAA name & Commits associated with the FBAA. \\
    \bottomrule
    \end{tabular}%
    }
  \label{tbl:repositoriesattributes}%
\end{table*}%

These repositories and their unique artifact ids (e.g., requirement id, patch id, and contribution id) allowed us to trace the contributions and commits to their architectural assets, product features, and platforms, via the patches that developers create and commits to internal source code branches. Table ~\ref{tbl:repositoriesattributes} presents the repositories including their attributes.

\textbf{The product portfolio repository} is used to support Sony Mobile's software platform strategy, where one platform is reused across multiple phones. The repository stores the different configurations between platforms, hardware and other major components along with market and customer related information. 
\textbf{The feature repository} stores information about each feature, which can be assigned to and updated by different roles as the feature passes through the firm's product development process. Information saved includes documentation of the feature description and justification, decision process, architectural notes, impact analysis, involved parties, and current implementation state. The \textit{contribution strategy} attribute is used to communicate the decisions from the CAP model usage, on whether the feature should be contributed or not.

\textbf{Feature-Based Architectural Asset (FBAA) repository} (FBAAs) groups features together that make up common functionality that can be found in multiple products, e.g. features connected to power functionality may be grouped together in its own FBAA and revised with new versions as the underlying features evolve along with new products. Products are defined by composing different FBAAs which can be considered as a form of configuration management. 

Even though Sony Mobile uses Android as an underlying platform, customization and new development are needed in order to meet customers' expectations. These adaptations are stored as patch artifacts in the \textbf{patch repository}. The patch artifacts contain information about the technical implementation and serve as an abstraction layer for the code commits which are stored in a separate \textbf{commit repository}. Each patch artifact can be traced to both FBAAs and features.

The patches that are contributed back to the OSS ecosystems have associated  contribution artifacts stored in the \textbf{contribution repository}. These artifacts store information such as the type of contribution and complexity, responsible manager and contributor, and concerned OSS ecosystem. Each contribution artifact can be traced to its related patch artifact.

With this set-up of repositories and their respective artifacts, Sony Mobile can gather information necessary to follow up on what functionality is given back to OSS ecosystems. Moreover, Sony Mobile can also measure how much resources that are invested in the work surrounding the implementation and contribution. Hence, this set-up makes up a critical part in both the structuring and execution of the CAP model.

This meta-model was created in the context of Sony Mobile's development organization. Hence, it is adapted to fit Sony Mobile' software product line strategy with platforms from which they draw their different products from. The architectural assets (FBAAs) play a key part in this configuration management. As highlighted in section~\ref{sec:Externalvalidity}, we believe that the meta-model will fit organizations in similar characteristics, and for other cases provide inspiration and guidance. This is something that we aim to explore and validate beyond Sony Mobile in future design cycles.

\section{Combining the CAP Model and the Information Meta-model}
\label{sec:Example}
In this section, we provide an example of how the CAP model may be used to classify an artifact, and combine this with the information meta-model to support communication and follow-up of the artifact and its decided contribution strategy. The example is \textit{fictive}\footnote{Due to confidentiality reasons, we have to select this example.} and was derived together with one of the experts (I2) from Sony Mobile with the intention to demonstrate the reasoning behind the artifact classification. Following the proactive process defined in section~\ref{subsec:Proactive}, we begin by discussing scope and abstraction level. 

For Sony Mobile, FBAAs offer a suitable abstraction level to determine whether certain functionality (e.g., a media player or power saving functionality) can be contributed or not. If the artifact is too fine-grained it may be hard to quantify its business impact and control complexity. In these cases, features included in a certain FBAA would inherit the decision of whether it can be contributed or not. Regarding the scope, we look at FBAAs related to the telephony part of a certain platform-range. The FBAA that we classify regards the support for Voice over Long-Term Evolution (VoLTE), which is a standard for voice service in the LTE mobile radio system~\cite{poikselka2012voice}. Note that this classification is performed when VoLTE was relatively new to the market in 2015.

VoLTE is classified in regard to its business impact and control complexity. The questions defined in section~\ref{subsec:Proactive} were used. Under each question, we provide a quote from I2 about how (s)he reasons, and the score which can be in the range of 1-4. We start by addressing the business impact:

\begin{enumerate}
    \item How does it impact on the firm's profit and revenue? \\
    \textit{``VoLTE is hot and an enabler for services and the European operators are very eager to get this included. This directly affects the firm's ability to range its products at the operators. So very important. Is it super important? The consumers will not understand the difference of it, they will get it either way.''} - \textbf{Score: 3}.
    \item How does it impact on the customer and end user value? \\
    \textit{``The consumers themselves may not know about VoLTE, but they will appreciate that the sound is better and clearer because other coding standards may be used.''} - \textbf{Score: 3}.
    \item How does it impact on the product differentiation? \\
    \textit{``VoLTE has a positive effect. Some product vendors will have VoLTE enabled and some not. So there is a differentiation which is positive. Does this have a decisive effect concerning differentiation? Is it something that the consumers will interpret as something that is very important? No.''} - \textbf{Score: 3}.
    \item How does it impact on the access to leading technology/trends? \\
    \textit{``VoLTE is very hot and is definitely a leading technology.''} - \textbf{Score: 3}.
    \item How does it impact if there are difficulties or shortages? \\
    \textit{``If we cannot deliver VoLTE to our customers, how will that affect them? It will not be interpreted as positive, and will not pass us by. But they will not be fanatic about it.''} - \textbf{Score: 2}.
\end{enumerate}

This gives us a mean score of 2,8. We repeat the same process for control complexity:

\begin{enumerate}
    \item Do we have knowledge and capacity to absorb the technology? \\
    \textit{``Yes, we have. We are not world experts but we do have good knowledge about it.''} - \textbf{Score: 3}.    
    \item Are there technology availability barriers and IPR constraints? \\
    \textit{``Yes, there were some, but not devastating. There are patents so it is not straight forward.''} - \textbf{Score: 2}.
    \item What is the level of innovativeness and novelty? \\
    \textit{``It is not something fantastic but good.''} - \textbf{Score: 3}.
    \item Is there a lack of alternatives? \\
    \textit{``Yes, there are not that many who have development on it so there are quite a few options. So we implemented a stack ourselves.''} - \textbf{Score: 3}.
    \item Are there limitations or constraints by the firm? \\
    \textit{``No, there are none. There is not a demand that we should have or need to have control over.''} - \textbf{Score: 1}.
\end{enumerate}

\begin{figure*}
\begin{center}
\includegraphics[width=\textwidth]{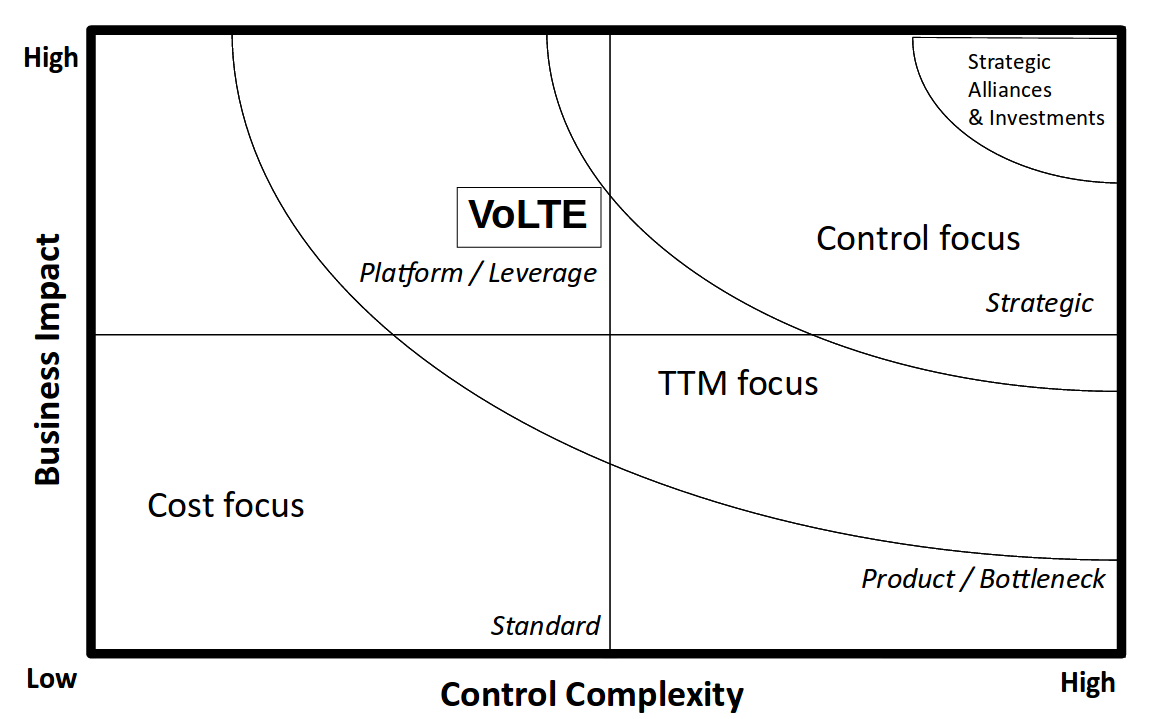}
\label{fig:VoLTE_example}
\caption{The CAP model and the example of VoLTE which is classified in regard to its business impact and control complexity.}
\end{center}
\end{figure*}

This gives us a mean score of 2,4. This places VoLTE in the bottom between of the upper two quadrants; the strategic and platform/leverage artifact quadrants. I2 elaborates on the strategy chosen: 

\textit{``VoLTE is an opportunity for us. We should invest in this technology, but we do not have to develop our own solution. Rather, we should take what is available externally. We should do active contributions, not just to get rid of maintenance, but also to push the technology forward with a time-to-market as our main contribution objective. It does not matter if it is open source. This is not rocket science that only we know about. We should have an open attitude towards VoLTE and support it as OSS and invest in it.''}

After reiterations and discussions, the decisions should be documented and communicated to the development organization. In Sony Mobile's case, the information meta-model is already integrated into requirements and product management infrastructure. Thus, these decisions would be added to the contribution strategy attribute of the feature artifacts which belong to the VoLTE FBAA artifact. To monitor and follow-up on the contribution strategy execution for VoLTE, product management can trace patch artifacts connected to the VoLTE feature artifacts, and see which of these that have contribution artifacts connected to them.


\section{Case Studies}
\label{sec:CaseStudies}
To perform a first validation of the CAP model outside Sony Mobile, we have conducted three exploratory case studies where we applied the CAP model and investigated its applicability and usability. Further and more extensive application and validation are planned for future design cycles. Below we present the results from this validation per case firm, which due to confidentiality reasons are made anonymous and referred to as firm A-C. For each firm, we present general characteristics, and how we conducted the case study. We then give a brief overview of their overall contribution strategy, followed a summary of the application of the CAP model, and an evaluation of the model in terms of its usability. For an overview, see table~\ref{tbl:casefirms}.

\begin{table}[]
\centering
\caption{Overview of the three case firms in regard to their domain, use of OSS, scope and abstraction analyzed with the CAP, and the setting in which the model was applied.}
\label{tbl:casefirms}
\begin{tabular}{p{1.1cm}p {3cm} p{1.8cm} p{2cm} p{2cm}}
\toprule
 & \textbf{Description} & \textbf{Use of OSS} & \textbf{Scope \& Abstraction} & \textbf{Setting} \\ \midrule
\textbf{Firm A} & Small-sized firm building a platform product for the agricultural domain. & OSS components in platform. & Features in platform product. & Interview with CTO. \\
\textbf{Firm B} & Small-sized firm building mobile games for mobile platforms. & OSS components in game products. & Features in a specific game. & Interview with Founder. \\
\textbf{Firm C} & Large-sized firm in the telecommunication domain. & OSS in service infrastructure. & Internal infrastructure project & Workshop with 8 cross-functional participants. \\ \bottomrule
\end{tabular}
\end{table}

\subsection{Case Firm A}
Firm A operates in the agriculture business. The main product of the firm is software designed to improve the efficiency of global grain marketing. The software offers a communication platform between the growers and buyers combined with real-time market intelligence. The main benefit is an enhanced ability to quickly respond to domestic and global market demands. We interviewed the CTO of the firm who has over 25 years of experience in the IT sector and was involved in 10 start-ups and many projects. 
The CAP model was used to analyze the current product the firm is offering. 

\subsubsection{Overall Contribution Strategy}
The firm makes extensive use of OSS code as long as it is not released under the GPL version 3 license. The firm keeps its own copy of the source code and often contributes bug fixes or other small changes, however without following up if they are integrated into the common code base. Decisions if to adapt the OSS ecosystem's version of the code are made on regular basis upon analysis. 

The firm has currently a static code policy that is based on the following reasoning. If the existing code works at the time, the firm does not care if it evolves and does not check if never versions are available. If there are changes, the firm checks first if the suggested improvements are beneficial before any new version is considered and integrated. 

Maintenance cost reduction is important for the firm, however not for the price of loosing competitive advantage. Thus, any functionality that has a differentiating potential is kept proprietary for about 6-9 months to check the market response and profitability. After this time, the firm analyzes if cost reduction is substantial before deciding to contribute the code or not. Estimating the current or future value of an asset is challenging, mainly because of rapid market changes and high market uncertainty. An example here is inventory management module that the firm's product has. This module (feature) turned out to be a strategic asset 12 months after developing it. So what may seem to be a rational decision from the development/technology perspective can be overwhelmed by market forces or conditions. Moreover, it may take a substantial amount of time before an intellectual property asset reveals its true value in the market place due to delays in the technology adoption curve. Therefore, cautious evaluation of the business and revenue values are necessary. If the technology adoption is slow, it is much more challenging and harder to see if and when to contribute. 

Regarding the contribution strategy, the firm has the following rules: 
\begin{itemize}
    \item high profit and critical to maintain control features are never shared with the OSS ecosystem as these build the firm's value in the eyes of the shareholders
    \item high profit and not critical to maintain control features - some resources are dedicated to investigate and see the potential of growing from low profit to high profit before a decision to contribute is made
    \item low profit and critical to maintain control features - the firm can release these features after commodity analysis. 
    \item low profit and not critical to maintain control features - the firm contributes these features as quickly as possible.  
\end{itemize}

The firm is small and in a growing phase with limited resources that can be dedicated to working with the OSS communities. The conclusion here is that OSS ecosystem engagement can be very valuable for large enterprises, in a resource constrained enterprise it is pretty risky policy.

\subsubsection{Application of the CAP Model}
Together with the firm's CTO, we have analyzed the current product with the help of the CAP model. The mapping of the product's features on the CAP model brings into focus the questions regarding: 1) where the differentiating value is, 2) what is the nature of the market the firm is operating in and 3) how much value the potential customers can absorb. This resulted in the following categorization: 

\begin{itemize}
    \item \textbf{Standard artifacts} - Covers about 20\% of all features. The CTO adds that not only OSS software is considered here but also binary modules. 
    \item \textbf{Product/Bottleneck artifacts} - Covers about 20\% of all features. These are mostly purchased or obtained from OSS communities to a lower time-to-market. An interesting aspect here is the usage of binaries that further reduces time-to-market as the integration time is lower compared to OSS modules that often require some scripting and integration efforts.
    \item \textbf{Strategic artifacts} - Covers only about 5\% of all features. The main reason is that the firm is afraid someone will standardize or control something in that part (interfaces) and destroy the shareholders' value. 
    \item \textbf{Platform/Leverage artifacts} - Covers about 55\% of all features because complexity is low and the firm has high control in case the firm becomes dominant in the market (they are currently not dominant). 
\end{itemize}

According to the CTO, a firm can be a "big winner" in immature markets that usually lack standards. Having a high portion of features in the Strategic artifact corner indicate operating in an established market where alliances need to be made do deliver substantial value. 

\subsubsection{Usability of the CAP Model}
The CTO indicated that the CAP model can be used by both executives and operational management. The primary stakeholder remains everyone who is responsible for product strategies. However, the executives will focus mostly on the strategy and if it reflects the direction given by the Board of Directors and main shareholders. In that regard, the percentage mapping of the features on the CAP model is considered useful as it shows where in those four quadrants (see Fig.~\ref{fig:OImodel}) a firm's product is, but also where it should be. When applied, there should be a cross-functional group as earlier suggested (see section~\ref{sec:Model}). The CTO agrees that a consensus-seeking approach should be used where opinions are first expressed independently, shared and then discussed until the group converges. This shows potential risks and additional uncovered aspects. 

When classifying artifacts in terms of business impact and control complexity, the CTO indicated that high-medium-low is sufficient in terms of scale. When several people perform the estimations, the results can show the density of each level for each aspect. The levels should be augmented with comments regarding additional risks or other important aspects. A scale of -1, 0 and 1 was also considered as suitable. 

The used frequency of the CAP model is estimated to be every major revision cycle when new features are added to the product. The complete analysis based on the CAP model should be performed when, e.g., entering the new market place or moving to more stable places in the market place. 


Our respondent believes that the CAP model usage delivers greater confidence that the firm is not deviating from the strategic direction and helps to identify the opportunities in the area in other quadrants. The usefulness was estimated as high and could be improved with more guidelines on how to interpret the mapping results. At the same time, it appears that larger organizations can benefit more from the CAP model application. The main problem for smaller firms with reaching high utility of the CAP model would be to have the resources to do regular analysis and the experience to provide valuable opinions. Experience in working with OSS and knowledge of the main driving forces for commoditization is considered essential. 

\subsection{Case Firm B}
Firm B develops mobile games for the Android and iOS platforms. 
The market place that the firm operates in is rather disordered and characterized by several players who use the same game engine that has a very active ecosystem\footnote{https://unity3d.com/} around it. A substantial part of the product is available for free with little integration effort. Reusing platforms and frameworks with large user base is an important survival aspect, regardless if they are OSS or not since acquisition costs are marginal. Entry barriers are negligible which implies that the commercial success is often a "hit and miss". In many aspects, the environment resembles an inverted OSS ecosystem where a given tool from a given provider or a given module is available with the source. Where a given tool or module from a given provider is available, often with source, at little or no charge. As a result, significant elements of the games are, essentially, commodities and product differentiation principally occurs within the media assets and the gameplay experience. The tool provider\footnote{https://unity3d.com/} is open sourcing back to the ecosystem and can gain those inverted benefits. The customers are helping the provider to improve the quality of the offering. The studied firm only report bugs to these ecosystems and never considers any active contributions or extensions. 

The mobile game users expect to play the game for free and perceive them as commodities. This impacts profitability and ability to be commercially viable. If the game is successful there are many opportunities to disturb the market place, e.g. a competitor copies the first 5 levels of the game and offers a similar copy to the market. About 80\% of the revenue is generated in the first five days after the game is released since the immediate customer behavior defines if the asset is worth something or not. 

\subsubsection{Overall Contribution Strategy} 
Since profitability decreases rapidly after product launch, firm B wants to directly minimize maintenance costs. This implies contributing the code base or using commodity parts as much as possible. Contribution strategy associated decisions need to be made rapidly based on the revenue trends and results. The odds of having long term playability for games other than adventure are very low. So for each release, the firm can receive a spike in the income and profitability and needs to carefully plan how to utilize this income. Time to market remains the main success factor in this market segment.  

Analyzing this market segment with the help of the CAP model brings forward how extreme the risk levels are in the mobile games business. CAP works well here as a risk assessment tool that should be applied to investments. In this market place, the quadrants of the CAP model can be merged and discussed together. The main analysis should be along the Y-axis and the discussion should be profit driven since the firm does not have any control over the platform, but controls the player experience.

Regarding the contribution strategy, the firm has the following rules: 
\begin{itemize}

\item high profit and critical to maintain control features - these features are considered as key differentiators but in this context there are very low barriers to copying by fast followers that clone the features. So keeping the features proprietary does not eliminate the risk of "fast clones". 
\item low profit and not critical to maintain control features - firm B obtains these features from 3rd party suppliers. 
\item low profit and not critical to maintain control features - firm B tried to obtain the components from 3rd parties and if it is not possible the software architecture is changed to eliminate criticality. 
\item high profit and not critical to maintain control features - there are no features with this characteristics according to firm B. 

\end{itemize}

\subsubsection{Application of the CAP Model}
We mapped the product features to the CAP model grid. The results are: 0\% of the features in the low left quadrant (\textbf{Standard artifacts}), 15\% in low right quadrant (\textbf{Product/Bottleneck artifacts}), 80\% in upper left quadrant (\textbf{Platform/Leverage artifacts}) and 5\% in top right (\textbf{Strategic artifacts}). Because firm B works cross platform they are dependent on the platform provider and obtain other modules from the ecosystem, e.g. the 2d elements and the networking elements. Firm B hopes that remaining focused on the upper left corner is sufficient to get some customers. The firm is "at the mercy of" the other firms dominating the top right corner. CAP helps to points out here that the vast bulk of the technology that enables the experience is already a commodity and freely available so the only differentiating side is the game experience, but this is substantial investment in media, marketing, UI, graphics, and art-work.

\subsubsection{Usability of the CAP Model}
The CAP model helps to raise attention that the market is very competitive. The commodity price is very low, differentiation is difficult and acquisition costs are marginal. For firm B, it means that it is cheaper to pay someone else for development than to participate in OSS migration and integration. The main benefit from CAP application remains the conclusion that in mobile game development the focus needs to be on business impact. It is important to perform extensive analysis on the Y-axis for checking if a future game is commercially viable before analyzing the complexity dimension. 

The CAP model, in this case, can be used once and the clear conclusion for the firm is that it should change its market focus. The model clearly points out that if a firm is relatively new to mobile game development there is little profitability in this market unless you have 20-30 million dollars to invest in marketing and other actions to sustain long terms revenues. Our respondent believes that every new game concept can be and should be evaluated with the help of the CAP model.

Our respondent believes that the questions in the CAP model should be answered with the high, medium and low scale during a consensus-driven discussion. Since most of the discussion in on the Y-axis, the simple 3-point scale was considered sufficient. Our respondent also pointed out that the CAP model could potentially be extended to include hedonic qualities since a firm sells experience rather than software applications. 
Investing in a high complex game is very risky so firms in this domain tend to stay away from high complexity endeavors that are risky.

\subsection{Case Firm C}
Firm C operates in the telecommunication domain and extensively uses OSS to deliver software products and services. We applied the CAP model on one of the internal software infrastructure projects with the objective to support the decision process in regard to whether the project should be released as OSS. CAP was therefore used on a project level, instead of a set of features. We invited 8 participants from various functions at the firm (open source strategy, community management, legal, product management, and development) into a workshop session where the CAP model was discussed and applied.

\subsubsection{Overall Contribution Strategy}
Decisions on what projects that are released as OSS and what may be contributed to existing OSS projects are made by the OSS governance board, similar to that of Sony Mobile (see section~\ref{subsec:Reactive}). The board is cross-functional and includes the representatives from OSS strategy, legal, technology and software development.

Contribution requests are submitted to the OSS governance board from the engineering teams and usually concern projects related to the development tool-chain or the infrastructure technology stack. The requests are usually accepted given that no security threats are visible or potential patents can be disclosed. In addition, the board analyses the potential for creating an OSS ecosystem around the project to be released.

\subsubsection{Application of the CAP Model}
The studied project was first discussed in terms of its background and functionality in order to synchronize the knowledge level among the workshop participants. This was followed by a discussion of the project's business impact. The questions outlined in Section~\ref{subsec:Proactive} were used for framing the discussion, but instead of using the Likert scale of 1-4, the workshop participants opted for an open consensus-seeking discussion from start. 

The workshop participants agreed that the project has a high impact in terms profit and revenue, as it increases operational efficiency, decreases the license-costs, and increases security. As it is an internal infrastructure project used to deliver software products and services, it has limited impact on the customers and end-users. The technology is not seen as differentiating towards competitors but does enable easier access to new technology-standards that may have a substantial impact on the business. The firm's engineering department has managed to perform the daily operations and deliver the firm’s services without the use of the project, why it would not devastate business if it was no longer available. However, it does offer clear advantages which would cause a negative impact if it the availability was reduced or removed.

In regard to control complexity, it was concluded that the firm has the competence needed to continue developing the project. Further, the project did not include any IP and patents from the firm’s defensive patent portfolio. The underlying knowledge and technology can be considered as commodity. However, there is a lack of alternates as only two could be identified, both with shortcomings. Internally of the firm, there is a defined need for the project, and that influence on its development is needed. There is, however, no demand that the firm should maintain absolute control, or act as an orchestrator for the project.

The workshop participants classified the project as a strategic artifact due to the high business impact, as well as a relative need for control and lack of alternatives. Due to the latter reasons, the project should be released as a new OSS ecosystem as soon as possible in order to maintain the first-mover-advantage and avoid having to adapt to competing solutions. Hence, the main contribution objective should be to reduce time-to-market. The participants stated that the goal would be to push the project towards commodity, where the main objective would be to share the maintenance efforts with the ecosystem and refocus resources on more value-creating activities.

\subsubsection{Usability of the CAP Model}
The workshop participants found that the CAP model provided a useful lens through which their OSS governance board could look at contribution requests and strategically plan decisions. One participant expressed that the CAP model offers a blue-print to identify what projects that are more important to the firm, and align contribution decisions with internal business and product strategies by explicitly considering the dimensions of business impact and control complexity.

The workshop-participants preferred the open consensus-seeking discussions as a mean to determine the business impact and control complexity, and based on this classify the artifact to the most relevant artifact type and contribution strategy. The chosen strategy and aligning contribution objective could then be used to add further depth and understanding to the discussion, which helped the group to arrive at a common decision and final contribution strategy for the reviewed project.

The questions defined in section~\ref{subsec:Proactive} were found useful to frame the discussions. Participants expressed that these could be further customized to a firm, but that this should be an iterative process as the OSS governance board applies the CAP model when reviewing new projects. The participants further expressed that some questions are more relevant to discuss for certain projects than others, but they provide a checklist to walk through when reviewing a project.

\section{Discussion}
\label{sec:Discussion}
In this section, we discuss the applicability and usability of the CAP model. We discuss the findings from the case studies how the CAP model should be improved or adapted to fit other contexts.  

\subsection{Applicability and Usability of the CAP Model}
The three cases presented in section~\ref{sec:CaseStudies} bring supporting evidence that the CAP model can be applied on: a set of features, a product or on a complete project. The model has proven to bring useful insights in analyzing a set of features in a product with the indication that larger organizations can benefit more from the CAP application than small organizations. In case B, the application of CAP provided valuable insights regarding the nature of the market and the risks associated with making substantial investment in this market. In case C, the application of the CAP model provide a lens though which the OSS governance board can screen current projects and decide upon their contribution or OSS release strategies. 

CAP was found useful as decision-support for individuals, executives and managers. However, as highlighted by respondents from firms A, B and C, CAP is best suited for a cross-functional group where consensus-seeking discussions can be used to bring further facets to the discussions and better answer the many questions that needs to be addressed. As for Sony Mobile and case firm C, a suitable forum for large-sized firms would be the OSS governance boards or OSS program offices.

The questions suggested in section~\ref{subsec:Proactive} were found useful, but it was highlighted that these may need to be tailored and extended as CAP is applied to new projects and features. When answering the questions and determining the dimensions of business impact and control complexity, the cases further showed that on scale does not fit all. Case firm A and B suggested a high-medium-low scale, while case firm C preferred to use the consensus-seeking discussion with out the help of a scale. These facts highlight that certain adaptations are needed for the CAP model to maintain its usability and applicability in different settings. It also highlights that the decision process should not be "over-engineered". Our results suggest that complexity needs to be balanced in order to maintain usability for the practitioners while still keeping the applicability on different types of artifacts and settings. How to adapt this balancing act and tailor the CAP model to different settings is a topic for future design cycles and case evaluations.

\subsection{Influence Needed to Control}
The Kraljic's portfolio model was originally used to help firms to procure or source supply-items for their product manufacturing~\cite{kraljic1983purchasing}. One of the model's two decision factors is \textit{supply risk}. To secure access to critical resources, a certain level of control is needed, e.g., having an influence on the suppliers to control the quality and future development of the supply-items. For OSS ecosystems, this translates into software engineering process control, for example in terms of how requirements and features are specified, prioritized and implemented, with the goal to have them aligned with the firm's internal product strategy. 

Software artifacts with a high control complexity (e.g., the media frameworks for Sony Mobile, see section~\ref{subsec:Strategic Artifacts}) may require special ownership control and a high level of influence in the concerned OSS ecosystems may be warranted to be able to contribute them. In cases where a firm does not posses the necessary influence, nor wish to invest the contributions and increased OSS activity~\cite{dahlander2005relationships} which may be required, an alternative strategy is to share the artifact with a smaller set of actors with similar agendas, which could include direct competitors~\cite{west2006challenges}. This strategy is still in-line with the meritocracy principle as it increases the potential ecosystem influence via contributions~\cite{dahlander2005relationships}. Sharing artifacts with a limited number of ecosystem actors leaves some degree of control and lowers the maintenance cost via shared ownership~\cite{stuermer2009extending, ven2008challenges}. Further, time-to-market for all actors that received the new artifacts is substantially shortened. 

For artifacts with less complexity control, e.g., those concerning requirements shared between a majority of the actors in the OSS ecosystem, the need for control may not be as high, e.g., the DLNA project or Linux commodity parts, see sections~\ref{subsec:Platform/Leverage Artifacts} and~\ref{subsec:Standard Artifacts}. In these cases, it is therefore not motivated to limit control to a smaller set of actors which may require extra effort compared to contributing it to all ecosystem actors. An alternative implementation may already be present or suggested which conflicts with the focal firm's solution. Hence, these types of contributions require careful and long term planning where the influence in the ecosystem needs to be leveraged. In case of firm B, complexity is controlled by the framework provider. 

For both critical or less critical artifacts in regard to control complexity, a firm needs to determine the level of influence in the involved ecosystems. This factor is not explicitly covered by the CAP model and could be considered as an additional discussion point or as a separate decision factor in the contribution strategies which are elicited from the CAP model.

\subsection{Direct and Indirect Use of OSS ecosystems}
The second decision factor originating from the Kraljic's model~\cite{kraljic1983purchasing} is the \textit{profit impact}. Profit generally refers to the margin between what the customer is willing to pay for the final product and what the product costs to produce. For OSS ecosystems, this translates into how much \textit{value} a firm can offer based on the OSS, e.g. services, and how much resources the firm needs to invest into integration and differentiation activities. I.e., much of the original definitions are preserved in the CAP model and the re-labeled decision factor business impact.

Artifacts with high profit, or high business impact are differential towards competitors and add significant value to the product and service offerings of the firm~\cite{van2009commodification}, e.g., the gaming services for Sony Mobile, see section~\ref{subsec:Strategic Artifacts}. Analogous, artifacts with low profit are those related to commodity artifacts shared among the competitors, e.g., Linux commodity parts, see section~\ref{subsec:Standard Artifacts}. This reasoning works in cases where the OSS and its ecosystem is directly involved in the product or service which focal firm offers to its customers. The customers are those who decide which product to purchase, and therefore mainly contribute in the value creation process~\cite{aurum2007value}. This requires good customer-understanding to judge which artifacts are the potential differentiators that will influence the purchase decision. 

In cases where an OSS has an indirect relation to the product or service of the firm, the artifact's value becomes harder to judge. This is because the artifact may no longer have a clear connection to a requirement which has been elicited from a customer who is willing to pay for it. In these cases, firms need to decide themselves if a particular artifact gives them an advantage relative to its competitors. 

OSS ecosystems often facilitates software engineering process innovations that later spark product innovations that increase the business impact of an artifact, e.g., if an artifact makes the development or delivery of the product to a higher quality or shorter time-to-market respectively~\cite{linaaker2015survey}. These factors cannot be judged by marketing, but rather by the developers, architects and product managers who are involved on the technical aspects of software development and delivery. In regard to the CAP model, this indirect view of business impact may be managed by having a cross-functional mix of internal stakeholders and subject-matter experts that can help to give a complete picture of an artifact's business impact.

\subsection{Comparing to Other Commoditization Models}
Both commoditization models suggested by van der Linden et al.~\cite{van2009commodification} and Bosch~\cite{Bosch13} consider how an artifact moves from a differential to a commoditized state. This is natural as technology and functionality matures and becomes standardized among actors on the same market or within the same OSS ecosystem. The impact of whether an artifact is to be considered differential or commodity is covered by the business impact factor of the CAP model. However, how quickly an artifact moves from one state to another is not explicitly captured by the CAP model. This dimension requires firms to continuously use the CAP model and track the evolution of features and their business impact. We recommend that the evaluation is performed every time a new product is planned and use the reactive approach in combination with the proactive (see section~\ref{subsec:Reactive} and~\ref{subsec:Proactive} respectively).

Relative to the level of commoditization of an artifact, the two previous models consider how the artifact should be developed and shared. Van der Linden et al.~\cite{van2009commodification} suggested to internally keep the differential artifacts and gradually share them as they become commoditized through intra-organizational collaborations and finally as OSS. In the CAP model, this aligns with the control complexity factor, i.e., how much control and influence is needed in regard to the artifact. 

The main novelty of the CAP model in relation to the other commoditization models~\cite{van2009commodification, Bosch13} considers OSS ecosystem participation and enables improved synchronization towards firms' product strategy and product planning, via feature selection, prioritization and finally release planning~\cite{kittlaus2008software}. The strategic aspect covered by the CAP model uses the commoditization principle together with business impact estimates and control complexity help may firms to better benefit from potential OI benefits. Assuming the commoditization is inevitable, the CAP model helps firms to fully benefit the business potential of differential features and timely share them with OSS ecosystems for achieving lower maintenance costs. Moreover, the CAP model helps to visualize the long term consequences of keeping or contributing an internally developed software artifact (more patches and longer time-to-market as consequence). Finally, the CAP model provides guidelines for how to position in an OSS ecosystem's governance structure~\cite{baars2012framework} and how to influence it~\cite{dahlander2005relationships}.

There may be various reasons why a firm would wish to contribute an artifact. Thus, the drivers used by Sony Mobile in the CAP model may not be the same for other firms wishing to adopt the model. The identified contribution drivers and cost structures should be aligned with the firm's understanding for how the value is drawn from the OSS ecosystems. This may help to improve the understanding of what should be contributed and how the resources should be planned in relation to these contributions. How the contribution objectives and drivers for contributions needs to be adapted is a topic for future research.

\section{Conclusion}
\label{sec:Conclusions}

The recent changes in software business have forced software-intensive firms to rethink and re-plan the ways of creating and sustaining competitive advantage. The advent of OSS ecosystems has accelerated value creation, shortened time-to-market and reshaped commoditization processes. Harvesting these potential benefits requires improved support for strategic product planning in terms of clear guidelines of \textit{what to develop internally} and \textit{what to open up}. Currently available commoditization models~\cite{van2009commodification,Bosch13} accurately capture the inevitability of commoditization in software business, but lack operational support that can be used to decide \textit{what} and \textit{when} to contribute to OSS ecosystems. Moreover, the existing software engineering literature lacks operational guidelines, for how software-intensive firms can formulate contribution strategies for improved \textit{strategic product planning} at an artifact's level (e.g., features, requirements, test cases, frameworks or other enablers).   

This paper introduces the Contribution Acceptance Process (CAP) which is developed to bridge product strategy with operational product planning and feature definition (\textbf{RQ1}). Moreover, the model is designed with commoditization in mind as it helps in setting contribution strategies in relation to the business value and control complexity aspects. Setting contribution strategies allow for \textit{strategic product planning} that goes beyond feature definition, realization and release planning. The CAP model was developed in close collaboration with Sony Mobile that is actively involved in numerous OSS ecosystems. The model is an important step for firms that use these ecosystems in their product development and want to increase their OI benefits, such as increased innovation and shorter time-to-market. This paper also delivers an information meta-model that instantiates the CAP model and improves the communication and follow-up of current contribution strategies between the different parts of a firm, such as management, and development (\textbf{RQ2}). 

There are several important avenues for future work around the CAP model. Firstly, we aim to validate the CAP model and related information meta-model in other firms, both statically and dynamically. We plan to focus on understanding the firm specific and independent parts of the CAP model. 
Secondly, we plan to continue to capture operational data from Sony Mobile and the three case firms related to the usage of the CAP model that will help in future improvements and adjustments. 
Thirdly, we plan to investigate how a contribution strategy can consider the influence a firm needs in an OSS ecosystems to be able to exercise control and introduce new features as needed. We believe that gaining and maintaining such influence in the right ecosystems is pivotal in order to execute successfully on contribution strategies.
Fourthly, we want to investigate to what degree the CAP model supports innovation assessment for firms not working with OSS ecosystems. Our assumption is that these firms could use the CAP model to estimate the degree of innovativeness of the features (could be considered as an innovation benchmark) without setting contribution strategies. 
Lastly, we plan to explore which technical aspects should be considered and combined with the current strong business view of the CAP model (e.g. technical debt and architecture impact seems to be good candidates to be included). 

\section{Acknowledgement}
This work was funded by Vinnova in the ITEA2 project 12018 SCALARE, and supported by the IKNOWDM project [grant number 20150033] from the Knowledge Foundation in Sweden and the Swedish National Science Foundation Framework [grant number 621-2012-5354]. We also would like to thank Prof. Per Runeson and Prof. Bj\"{o}rn Regnell for valuable feedback and discussions regarding this study. Further, we would like to thank the four anonymous reviewers for their constructive and valuable comments provided in the review process.

\section*{References}
\bibliography{refs}

\end{document}